\documentclass[12pt,a4]{article}
\renewcommand{\baselinestretch}{1.25}

\usepackage{amsthm,amsmath,latexsym,amssymb,amsfonts,amscd}
\usepackage{graphics,fancyhdr,array,stmaryrd,euscript}
\usepackage{slashed}
\usepackage{diagrams}
\numberwithin{equation}{section}

\usepackage{hyperref}

  \usepackage[numbers,sort&compress]{natbib}
  \usepackage[nottoc]{tocbibind}

\newcommand{\non}{\nonumber}
\newcommand{\pl}{\partial}
\newcommand{\be}{\begin{equation}}
\newcommand{\ee}{\end{equation}}
\newcommand{\bea}{\setlength\arraycolsep{2pt} \begin{eqnarray}}
\newcommand{\eea}{\end{eqnarray}}
\newcommand{\eq}[1]{(\ref{#1})}
\newcommand{\w}[1]{\\[0.#1cm]}

\newcommand{\mm}{{\ensuremath{\underline{m}}}}
\newcommand{\nn}{{\ensuremath{\underline{n}}}}

\newcommand{\gad}{{\dot{\alpha}}}
\newcommand{\gbd}{{\dot{\beta}}}
\newcommand{\gdd}{{\dot{\gamma}}}

\newcommand{\ga}{\alpha}
\newcommand{\gb}{\beta}
\newcommand{\gc}{\gamma}
\newcommand{\gd}{\delta}

\newcommand{\adD}{{\mathsf{D}}}
\newcommand{\tadD}{\widetilde{\mathsf{D}}}

\newcommand{\fd}[1]{{}^{\vphantom{#1}}_{#1}}
\newcommand{\fud}[2]{{}^{#1}{}_{#2}\,}
\newcommand{\fdu}[2]{{}_{#1}{}^{#2}\,}

\newcommand{\bry}{{{\bar{y}}}}

\newcommand{\bru}{{{\bar{u}}}}
\newcommand{\brv}{{{\bar{v}}}}

\newcommand{\brxi}{{{\bar{\xi}}}}

\newcommand{\Pib}{{\bar{\Pi}}}

\newcommand{\besubeqs}{\begin{subequations}}
\newcommand{\esubeqs}{\end{subequations}}

\newcommand{\omegatwo}{\omega^{(2)}}
\newcommand{\Ctwo}{C^{(2)}}
\newcommand{\omegaone}{\omega^{(1)}}
\newcommand{\Cone}{C^{(1)}}

\newcommand{\Fron}{{\Phi}}

\newcommand{\rmx}{{\mathrm{x}}}

\newarrow {Corresponds} <---> 

\begin{document}
%%%%%%%%%%%%%%%%%%%%%%%%%%%%%%%%%%%%%%%%%%%%%%%%%%%%%%%%%%%%%%%%%%%%%%%

\begin{flushright} 
MI-TH-17637\\
{LMU-ASC 27/17}
\end{flushright}
\vspace{10mm}

\begin{center}

{\Large {\bf Chern-Simons Matter Theories and Higher Spin Gravity}} \\[5mm]

\vspace{8mm}
\normalsize
{\large  Ergin Sezgin${}^{1}$, Evgeny D. Skvortsov${}^{2,3}$ and Yaodong Zhu${}^1$}

\vspace{10mm}
${}^1${\it Mitchell Institute for Fundamental Physics and Astronomy\\ Texas A\&M University
College Station, TX 77843, USA}
\vskip 1 em
${}^2${\it Arnold Sommerfeld Center for Theoretical Physics\\
Ludwig-Maximilians University Munich\\
Theresienstr. 37, D-80333 Munich, Germany}\\
\vskip 1 em
${}^{3}${\it  Lebedev Institute of Physics, \\
Leninsky ave. 53, 119991 Moscow, Russia}

%\vspace{20mm}
\vspace{15mm}

\hrule

\vspace{8mm}

\begin{tabular}{p{14cm}}
{\small
 We compute the parity violating three point amplitudes with one scalar leg in higher 
 spin gravity and compare results with those of Chern-Simons matter theories. 
 The three-point correlators of the free boson, free fermion, critical vector model 
 and Gross-Neveu model are reproduced including the dependence on the Chern-Simons coupling. 
 We also perform a simple test of the modified higher spin equations proposed 
 in \href{https://arxiv.org/abs/1605.02662}{arXiv:1605.02662 [hep-th]} and find that the 
 results are consistent with the AdS/CFT correspondence.
}
\end{tabular}
\vspace{7mm}
\hrule
\end{center}

\newpage

\setcounter{tocdepth}{2}
\renewcommand{\baselinestretch}{1.1}\normalsize
\tableofcontents
\renewcommand{\baselinestretch}{1.25}\normalsize

\newpage

%%%%%%%%%%%%%%%%%%%%%%%%%%%%%%%%%%%%%%%%%%%%%%%%%%%%%%%%%%%%%%%%%%%%%%%
\section{Introduction}
%%%%%%%%%%%%%%%%%%%%%%%%%%%%%%%%%%%%%%%%%%%%%%%%%%%%%%%%%%%%%%%%%%%%%%%

Higher spin (HS) theories provide simple models of AdS/CFT duality where the dual conformal 
field theories have matter in vector representation of the gauge group 
\cite{Klebanov:2002ja,Sezgin:2002rt,Sezgin:2003pt,Leigh:2003gk}. Such CFTs have a rather 
moderate growth of states and are usually exactly solvable in the large-$N$ limit. Moreover, 
free CFTs fall into this class too and are, in fact, generic duals of HS theories. Boundary 
values of HS gauge fields are sources of conserved tensors whose presence is a distinguished 
feature of free CFTs in $d\geq3$ \cite{Maldacena:2011jn,Boulanger:2013zza,Alba:2013yda,Alba:2015upa}. 
They also occur in the strict $N=\infty$ limit of interacting vector models. In this paper we 
study a rich class of $AdS_4/CFT_3$ dualities between Chern-Simons matter theories 
\cite{Giombi:2011kc,Aharony:2011jz} and four-dimensional HS theories \cite{Giombi:2011kc}. 
Free theories are usually isolated as well as vector models, but it turns out that in $3d$ 
there is a one-parameter family of theories that are defined by coupling free or interacting 
vector models to Chern-Simons gauge field, which does not introduce any new local operators 
and does not break conformal symmetry. Parity is broken in Chern-Simons matter theories and 
correlation functions of HS currents have a very special structure \cite{Maldacena:2012sf}: they 
consist of several terms, some of which occur in free theories and some other break parity. 
In \cite{Giombi:2012ms} these parity-breaking terms were found by direct computation in Chern-Simons 
matter theories. In this paper we reproduce some of these structure from the $AdS_4$ side by 
computing the three-point amplitudes involving a scalar field.
The subsector of HS theory needed for the computation of this amplitude is distinguished by the fact that the interactions 
are local and are fixed by the HS algebra structure constants and does not require dealing with 
non-localities present in the Vasiliev equations. 

\begin{figure}[h]
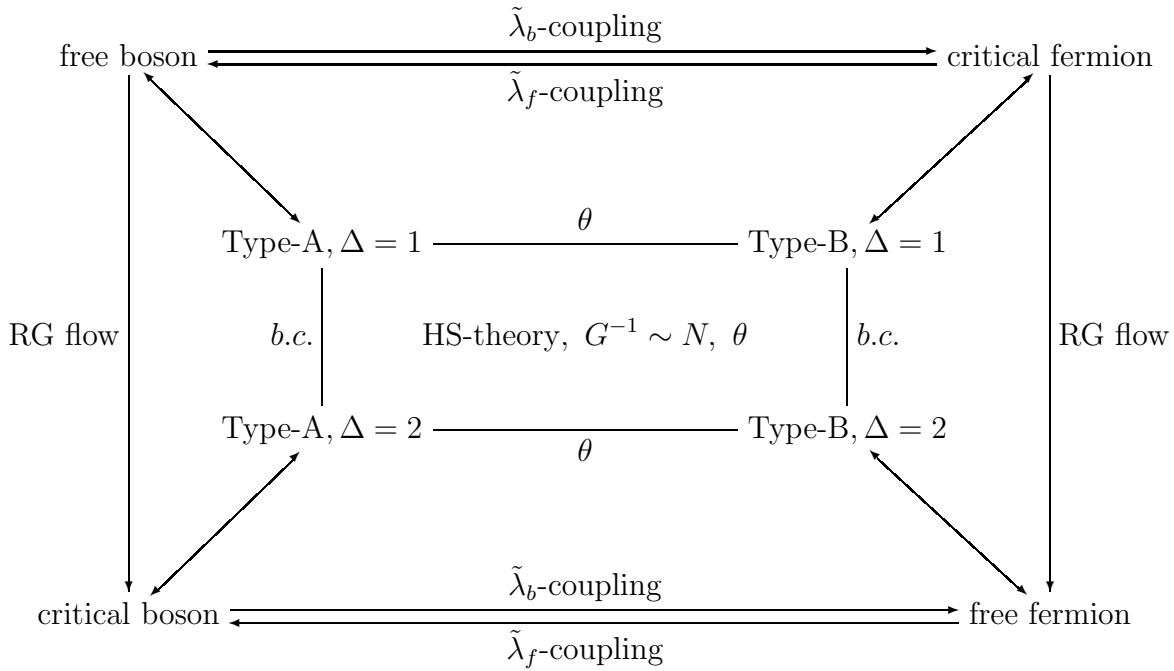

\begin{diagram}
\text{free boson} & && \pile{\rTo(6,0)^{\text{$\tilde{\lambda}_b$-coupling}} \\ \lTo_{\text{$\tilde{\lambda}_f$-coupling}}} &&&  
\text{critical fermion}\\
& \rdCorresponds & & & & \ldCorresponds&\\
& & \text{Type-A}, \Delta=1 & \rLine^{\theta} & \text{Type-B}, \Delta=1 & &\\
\dTo^{\text{RG flow}}& &\dLine^{b.c.} & \text{HS-theory},\  G^{-1}\sim N,\ \theta & \dLine_{b.c.} & &\dTo_{\text{RG flow}}\\
& & \text{Type-A}, \Delta=2 & \rLine_{\theta} & \text{Type-B}, \Delta=2 & & \\
& \ldCorresponds & & && \rdCorresponds&\\
\text{critical boson} & && \pile{\rTo(6,0)^{\text{$\tilde{\lambda}_b$-coupling}} \\ \lTo_{\text{$\tilde{\lambda}_f$-coupling}}} &&&  
\text{free fermion}
\end{diagram}
\caption{Schematic depiction of the relationships between various CFTs in three dimensions and 
higher spin gravity in four dimensions. 
The latter has two coupling constants, namely $(G,\theta)$, and $\Delta=1$, $\Delta=2$ or 
mixed boundary conditions can be imposed on the scalar field. }
\end{figure}

The emerging relations between various CFTs in three dimensions and HS theories in four dimensions 
are depicted in the diagram; see \cite{Giombi:2016ejx} for a review. In the upper left corner is 
free bosonic CFT, which RG flows to critical boson model under $(\phi^2)^2$ deformation. On the upper 
right is the critical fermion (Gross-Neveu) model with $(\bar\psi\psi)^2$ interactions, which RG flows 
to the free fermion CFT. The right arrow at the top denotes the coupling of level 
$k$ Chern-Simons theory and dialing of the 't Hoof coupling $\lambda_b=N/k$,\footnote{In practice, 
it is convenient to pass to effective $\tilde{N}$ and $\tilde{\lambda}$, which we will use throughout 
the paper. The relation to the microscopical parameters $N$ and $\lambda$ requires computation of 
some correlators, see \cite{Aharony:2012nh,GurAri:2012is}.} while the left arrow 
at the top denotes the coupling of CS theory to the critical fermion model and  dialing up the 
associated 't Hooft coupling $\lambda_f$. Similarly, the left arrow at the bottom 
denotes the coupling of CS theory to the free fermion model and dialing up the
't Hoof coupling $\lambda_f$, and the right arrow at the bottom denotes the coupling of CS theory 
to critical boson model and dialing up the 't Hooft coupling $\lambda_b$. The dialings of the 
effective 't Hooft couplings $\tilde{\lambda}$ are always from $0$ to $\infty$.

It turns out that the spectra of singlet operators coincide in  free fermion/critical boson and 
free boson/critical fermion pairs in the $N=\infty$ limit. This observation together with many 
other tests has led to the conjecture of three-dimensional bosonization 
\cite{Giombi:2011kc, Maldacena:2012sf, Aharony:2012nh, Aharony:2015mjs}: the bosonic theory at 
any $\lambda_b$ and $N_b$ can be equivalently described as the dual fermionic theory at some 
other $\lambda_f$ and $N_f$. More general dualities can be found  in \cite{Aharony:2015mjs,Seiberg:2016gmd,Karch:2016sxi}. 
The theory that  interpolates between the free fermion  and critical boson models is called quasi-fermionic 
(or CS-fermion) model. The theory that interpolates between free boson and critical fermion 
models is called quasi-bosonic (or CS-boson) model. Many tests of the three-dimensional bosonization 
duality have been already performed, see e.g. 
\cite{Aharony:2012nh,GurAri:2012is,Aharony:2012ns,Jain:2013py,Minwalla:2015sca}

AdS/CFT correspondence adds another direction to the diagram. Any of the four families of CFT's 
described above (there are actually only two families if the $3d$ bosonization conjecture is true) 
is expected to have an $AdS_4$ dual. By the field content, the natural candidate dual is a HS theory. 
Indeed, in the strict $N=\infty$ limit each of the theories possesses an infinite number of conserved 
tensors. The coupling constant in dual HS theory is of order $1/N$, while the t'Hooft constant 
$\lambda$ should be manifested as an additional coupling constant in the bulk theory. It is important 
to note that the Chern-Simons term breaks parity on the CFT side and therefore the dual HS theory 
should also violate parity unless we are at one of the limiting points. Remarkably, these are exactly 
the properties of the Vasiliev equations \cite{Vasiliev:1990vu,Vasiliev:1999ba}. There is a parameter 
that enters the equations in the form of a simple phase factor $e^{\pm i \theta}$. For generic $\theta$ 
the equations do not preserve parity. There are two parity preserving options: $\theta=0$ and 
$\theta=\pi/2$, which are called Type-A and Type-B \cite{Sezgin:2003pt}. On the $AdS_4$ side the 
difference between the duals of the CS-boson and CS-fermion models is in the boundary conditions for 
the scalar field of the HS multiplet. 
The relationships between the HS theories involved and their CFT duals are depicted in the diagram. 
The HS theories connected by the solid lines on the left (right) have $\theta=0$ $(\theta=\pi/2)$
and they differ from each other only with respect to the boundary conditions imposed on the scalar 
field present in HS theory. The upper (lower) solid lines denote the HS theory with generic $\theta$ 
parameter and  $\Delta=1 (\Delta=2)$ boundary conditions on the scalar field.

Since we are going to study the AdS/CFT three-point functions, let us review the results obtained so far. 
In \cite{Maldacena:2012sf} the effects of parity breaking were studied to the lowest order in $1/\tilde{N}$, 
but to all orders in $\tilde\lambda$. It was shown that the three point functions of 
HS currents $\langle j_s j_s j_s\rangle_{B,F}$ in
CS-boson and CS-fermion theories decompose into three structures: two of them represent the free theories 
('B' for CS-boson, 'F' for CS-fermion, 'b' for free boson and 'f' for free fermion) and another one 
is parity odd:
\begin{align}\label{MZA}
\langle jjj\rangle_{B,F}&={\tilde{N}}\left[ \cos^2 \theta \langle jjj\rangle_{b,f}
+\sin^2\theta \langle jjj\rangle_{f,b}+\cos\theta \sin\theta \langle jjj\rangle_{odd}\right]\,,
\end{align}
where $\cos^2\theta=1/(1+\tilde{\lambda}^2)$. At $\tilde{\lambda}=0$ the result simply states that 
the three-point functions are pure, i.e. are given by the free theory we started with.  As we switch 
on the coupling there is an admixture of the odd structure and, with a bit suppressed factor, the 
contribution of the structure of the opposite free theory (free fermion for CS-boson and free boson 
for CS-fermion). In the strongly coupled regime everything is upside-down: the contribution of the 
starting free theory is suppressed, there is a parity-odd piece and the leading contribution is due 
to the opposite free theory. In the limit of the strong coupling the three-point functions are pure, 
but are those of the opposite theory. This transition manifests the bosonization duality. Scalar singlet operators require special treatment and are not captured by \eqref{MZA}. In the first approximation the CS-boson theory has a scalar singlet operator $j_0$ of dimension $1$ and the CS-fermion has a scalar singlet operator $\tilde{j}_0$ of dimension $2$. Three-point functions with one insertion of $j_0$ ($\tilde{j}_0$) take a similar form given by \cite{Maldacena:2012sf} 
\begin{align}
\begin{aligned}
\langle j_{s_1} j_{s_2} j_0\rangle_{B.}&= {\tilde{N}} 
\left[\cos\theta \langle j_{s_1} j_{s_2} {j}_0\rangle_{f.b.}
+\sin \theta\langle j_{s_1} j_{s_2} j_0\rangle_{odd}\right]\ ,
\w2
\langle j_{s_1} j_{s_2} \tilde{j}_0\rangle_{F.}&= 
{ \tilde{N}} \left[\cos\theta \langle j_{s_1} j_{s_2} \tilde{j}_0\rangle_{f.f.}
+\sin \theta\langle j_{s_1} j_{s_2} \tilde{j}_0\rangle_{odd}\right]\ .
\end{aligned}
\end{align}
and we discuss them in Section \ref{subsec:csmatter}. The difference is that with each $j_0$ ($\tilde{j}_0$) insertion we lose one structure. Another remarkable fact is 
that the $\tilde{\lambda}$-deformation appears as an overall factor of some structures as displayed above, i.e. there 
are no spin-dependent factors. Moreover, it was argued in  \cite{Giombi:2011rz} and then proved in 
\cite{Giombi:2016zwa} that all the structures are unique. For example, there is just one parity-odd 
structure $\langle j_s j_s j_s\rangle_{odd}$. 

As far as the checks of the AdS/CFT duality are concerned, there exists a number of results. 
In \cite{Sezgin:2003pt} it was noted that the three-point coupling of the scalar field vanishes as 
obtained from the Vasiliev equation in the Schwinger-Fock gauge, which is consistent with vanishing 
of the three-point function in the critical vector model, corresponding to the $\Delta=2$ boundary condition on the bulk scalar. In the case of $\Delta=1$ boundary condition, however, a non-vanishing three point amplitude arises as a result of a regularization of a pertinent divergent integral \cite{Petkou:2003zz, Giombi:2009wh, Bekaert:2014cea}, in agreement with the result from the free boson model. In \cite{Giombi:2009wh} the first three-point functions 
were computed in the parity even case for two different spins $s_1$, $s_2$, the third one being 
zero (one scalar leg). It was assumed that the results give the correct CFT structures and it is 
possible to compare the leading coefficients only by choosing certain special kinematics. There are 
two types of terms in the HS equations that are relevant: we call them $\omega C$ (they are local) and 
$CC$ (may be non-local), which are explained in Section \ref{sec:hsinteractions}. The computation 
in \cite{Giombi:2009wh} was mostly based on $\omega C$ terms, which are local. We would like to point 
out that the $CC$-terms were found \cite{Giombi:2009wh} to produce either divergent results or, after 
some regularization, the results that are inconsistent with CFT. In \cite{Giombi:2012ms} it was noted 
that the $\omega C$-terms can produce parity-odd correlators. In particular, the $0-1-s$ parity-odd 
structure, which is fixed by conformal symmetry, was recovered in some limit.\footnote{See (5.8) there 
for the $AdS$ integral that is applicable for any $s_1,s_2$. }

In \cite{Giombi:2010vg} a different approach was adopted to compute amplitudes involving three 
different spins, which turns out to be equivalent, roughly speaking, to taking only the $CC$-terms 
into account.\footnote{The $\omega C$-terms seem to be missing, though there is a gauge choice ambiguity present in this method that could lead to such terms.} This approach exploits the extension of spacetime by spinorial coordinates. First of all, 
the $CC$-terms were found to yield a divergent result, which is consistent with the previous findings. 
Secondly, after some regularization in \cite{Giombi:2010vg}, the $CC$-terms give the parity-even 
correlators that are consistent with the duality. In fact, the regularization makes the $CC$-terms 
collapse onto the HS invariants, which are essentially the traces of the master fields in extended 
spacetime, or HS amplitudes \cite{Sezgin:2011hq,Colombo:2012jx,Didenko:2012tv}. These amplitudes are not represented as integrals of some vertex over the AdS space and for that reason they are not directly related to Witten diagrams. In fact, in this approach 
the amplitude computation amounts to a direct CFT computation without taking into account the local 
structure of AdS. However, this method, even if we accept the regularization, seems to lead to vanishing 
parity-odd correlators \cite{Giombi:2010vg}. 

Going back to the approach of \cite{Giombi:2009wh}, in  \cite{Boulanger:2015ova} the full structure 
of the second order corrections in the Vasiliev equations was worked out and it was shown that the 
resulting equations are too non-local to be treated by the field theory methods, which also explains 
infinities observed in \cite{Giombi:2009wh,Giombi:2010vg}. In \cite{Vasiliev:2016xui} a modification 
of some of the equations was proposed to cure the problem at the second order. 

In this paper we sharpen and expand considerably the tests of the HS holography discussed above. 
In particular we provide simplifications in the computation of HS correlation functions, and we 
establish nontrivial new tests of the holography for the parity violating HS theories. More specifically, 
our results can be summarized as follows. We isolate the scalar field equation in the Vasiliev theory up to 
quadratic terms in fields. This allows us to compute the $s_1-s_2-0$ correlators. The scalar field 
receives contribution only from $\omega C$-terms for $s_1\neq s_2$ and therefore does not rely on the 
structure of the $CC$-terms. The vertex is fixed by the HS algebra structure constants and also follows 
straightforwardly from the Vasiliev equations \cite{Vasiliev:1990en, Sezgin:2002ru}. We evaluate the 
vertex on the boundary-to-bulk propagators, which results in a rather simple expression. Next we compute 
the bulk integral and recover the full structure of the correlation functions for the Chern-Simons 
matter theories, including the four limiting cases, namely the free/critical fermion/boson theories. 
The dependence on the parity violating parameter comes out right. For the parity-even cases our results 
furnish an improvement of \cite{Giombi:2009wh} as the full structure is reproduced without assuming any 
special kinematics.\footnote{In \cite{Giombi:2009wh} a limit was taken such that the six conformally invariant structures that we review in the next Section get reduced to just one structure.} The parity-violating structures we find from the bulk computations are compared with 
recent results \cite{Giombi:2016zwa} for Chern-Simons matter theories and found to be in a perfect 
agreement. As a byproduct we find explicit formulae for the parity violating structures 
(in \cite{Giombi:2016zwa} they were encoded in a system of recurrence relations for the coefficients).

The correlators are analytic in spin and consequently the $s-s-0$ case can be reached too, 
as was already noted in \cite{Giombi:2009wh}, by evaluating the $s_1-s_2-0$ vertex for not equal spins and then setting $s_1=s_2$.\footnote{In \cite{Giombi:2012ms}, the $s-s-0$ bulk integral was shown to reproduce the generating function for even and odd correlators found in \cite{Maldacena:2011jn}. This generating function should give the right conformal structures whose normalization is yet to be fixed by comparing with CS-matter theories.} This gives non-zero contribution even though the $\omega C$-term vanishes identically for $s_1=s_2$ and the {\it full} $s-s-0$ amplitude originates from the $CC$-term alone. We perform the simplest test of the modified equations proposed in \cite{Vasiliev:2016xui} and recover the $s-s-0$ correlators 
with the right coefficient.

The outline of the paper is as follows. In Section \ref{sec:CFT} we review the structure of the CFT 
correlators in Chern-Simons matter theories. In Section \ref{sec:hsinteractions} we discuss the general 
structure of HS interactions and relation to the Vasiliev equations. In Section \ref{sec:Kinematic} we 
list the basic objects that constitute the boundary-to-bulk propagators for spinning fields. 
In Section \ref{sec:vertices} the propagators for HS fields are given, and the vertex is evaluated on 
the propagators. The actual computation of the cubic Witten diagram is done in Section \ref{sec:amplitude}. 
In Section \ref{sec:discussion}, we summarize and comment on our results, and we also discuss the contribution of the redefined $CC$ vertex \cite{Vasiliev:2016xui} to the $s-s-0$ amplitude. Our notations and conventions are given in Appendix \ref{app:notation}.

%%%%%%%%%%%%%%%%%%%%%%%%%%%%%%%%%%%%%%%%%%%%%%%%%%%%%%%%%%%%%%%%%%%%%%%
\section{CFT} 
\label{sec:CFT}
%%%%%%%%%%%%%%%%%%%%%%%%%%%%%%%%%%%%%%%%%%%%%%%%%%%%%%%%%%%%%%%%%%%%%%%

In this section we review the general structure of correlation function in three
dimensions and list the results available in the literature for free theories and, more generally, 
for Chern-Simons matter theories. The advantage of three dimensions is that one can benefit 
from the fact that $so(2,1)\sim sp(2)$. 

%%%%%%%%%%%%%%%%%%%%%%%%%%%%%%%%%%%%%%%%%%%%%%%%%%%%%%%%%%%%%%%%%%%%%%%
\subsection{General Structure of the Correlators} 
%%%%%%%%%%%%%%%%%%%%%%%%%%%%%%%%%%%%%%%%%%%%%%%%%%%%%%%%%%%%%%%%%%%%%%%

In three dimensions a traceless rank-$s$ $so(2,1)$-tensor is equivalent to a symmetric 
rank-$2s$ spin-tensor. It is convenient to contract the Lorentz indices of tensor operators 
with auxiliary polarization vectors that are now replaced by polarization spinors, which 
we denote by $\eta\equiv \eta^\ga$. Therefore, a weight-$\Delta$, rank-$s$ tensor operator 
$O^{a_1...a_s}_\Delta$ is replaced by a generating function $O_\Delta(\rmx,\eta)$:\footnote{$3d$ 
coordinates $\vec x$ are replaced by two-by-two symmetric matrices $\rmx^{\ga\gb}$. For further 
notations and conventions see Appendix \ref{app:notation}.}
\begin{align}
O^{a_1...a_s}_\Delta(\rmx) \qquad \longrightarrow \qquad O^{\alpha_1...\alpha_{2s}}_\Delta(\rmx^{\gb\gc}) 
\qquad \longrightarrow 
\qquad  
O_\Delta(\rmx,\eta)=O^{\alpha_1...\alpha_{2s}}_\Delta(\rmx)\eta_{\alpha_1}...\eta_{\alpha_{2s}}\ .
\end{align}
Suppose we are given a number of operators $O(\rmx_i,\eta_i)$ that are inserted at points 
$\rmx_i$ and whose tensor indices are contracted with polarization spinors $\eta^i_\ga$. 
The conformal group acts in the usual way. In particular, Lorentz transformations correspond 
to an $Sp(2)$ matrix $A\fdu{\ga}{\gb}$ that acts both on coordinates $\rmx_i$ and polarization 
spinors $\eta^i$:
\begin{align}
\rmx^{\gb\gd}&\rightarrow A\fdu{\ga}{\gb}A\fdu{\gc}{\gd}\rmx^{\ga\gc} \ ,
&& \eta^i_\ga \rightarrow A\fdu{\ga}{\gb} \eta^i_\gb\ .
\end{align}
It is useful to define the inversion map as 
\begin{align}
R\vec\rmx &=\frac{\vec \rmx}{\rmx^2}\ , & 
R\eta^a_\alpha &=\frac{\rmx_\alpha{}^\beta \eta_\beta^a} {\rmx^2}\ , & 
Rx^{\ga\gad}&=\frac{x^{\ga\gad}}{x^2} 
=\frac{\rmx^{\ga\gad}+iz\epsilon^{\ga\gad}}{\rmx^2+z^2}\ , 
\label{prime}
\end{align}
Then, it is not hard to see that the following structures are conformally invariant \cite{Giombi:2011rz}:
\begin{align}
P_{ij}&=  \eta^i_\ga R[\rmx_i-\rmx_j]^{\ga\gb}\eta^j_\gb\ , && RP_{ij}=P_{ij}\ ,
\\
Q^i_{jk}&= \eta^i_\ga \left(R[\rmx_j-\rmx_i]-R[\rmx_k-\rmx_i]\right)^{\ga\gb}\eta^i_\gb\ , 
&& RQ^i_{jk}=Q^i_{jk}\ .
\end{align}
There is also one more structure that is parity-odd:
\begin{align}
S^i_{jk} =  \frac{\eta^k_\ga (\rmx_{ki})\fud{\ga}{\gb}(\rmx_{ij})^{\gb\gc} \eta^j_\gc}{\rmx_{ij} \rmx_{ik} \rmx_{jk}}\ , 
&& RS^i_{jk}=-S^i_{jk}\ .
\end{align}
Any three-point correlation function 
$\langle O_1(\rmx_1,\eta^1)O_2(\rmx_2,\eta^2) O_3(\rmx_3,\eta^3)\rangle $ can be decomposed 
into an obvious prefactor times a polynomial in $Q,P,S$ structures:
\begin{align}
\langle O_1(\rmx_1,\eta^1)O_2(\rmx_2,\eta^2) O_3(\rmx_3,\eta^3)\rangle &= 
\frac{1}{\rmx_{12}^{\Delta_1+\Delta_2-\Delta_3} \rmx_{13}^{\Delta_1+\Delta_3
-\Delta_2}\rmx_{23}^{\Delta_2+\Delta_3-\Delta_1}}f(P,Q,S)\,.
\end{align}
The function $f$ must comply with the spin of the operators and also should not contain any 
redundant structures that are possible due to not all of $Q,P,S$ being independent. As we 
will need only two- and three-point correlation functions, it is convenient to introduce 
the following notation
\begin{align}
Q_1&\equiv Q^1_{32}\ , &Q_2&\equiv Q^2_{13}\ , & Q_3&\equiv Q^3_{21}\ , &
S_3&\equiv S^{3}_{21}\ .
\end{align}
The even power of any odd structure is even, which is manifested by \cite{Giombi:2011rz}
\begin{align}
S_3^2+Q_1 Q_2-P_{12}^2\equiv0\,.
\end{align}
This is the only relation we need for the $s_1-s_2-0$ correlators. The even structures $P,Q$ 
can be identified as the building blocks of the simplest correlators that are completely 
fixed by conformal symmetry
\begin{align}
\langle j_{s_1}(\rmx_1,\eta_1) j_{s_2}(\rmx_2,\eta_2) \rangle & 
\sim\frac{1}{\rmx^2_{12}} \delta_{s_1,s_2}(P_{12})^{s_1+s_2}\ ,
\w2
\langle j_{s_1}(\rmx_1,\eta_1) j_0(\rmx_2) j_0(\rmx_3)\rangle &
\sim\frac{1}{\rmx_{12}\rmx_{23}\rmx_{31}} (Q_1)^{s_1}\ ,
\end{align}
where we assumed that $j_s$ is the spin-$s$ conserved tensor and the weight of scalar 
operator $j_0$ is $\Delta=1$. The conservation of currents can be imposed with the help 
of a simple third order operator:
\begin{align}
\mathrm{div} = \frac{\pl^2}{\pl \eta_\ga  \pl \eta_\gb}\frac{\pl}{ \pl \rmx^{\ga\gb} }\ .
\end{align}

%%%%%%%%%%%%%%%%%%%%%%%%%%%%%%%%%%%%%%%%%%%%%%%%%%%%%%%%%%%%%%%%%%%%%%%
\subsection{Free Boson} 
\label{subsec:freeboson}
%%%%%%%%%%%%%%%%%%%%%%%%%%%%%%%%%%%%%%%%%%%%%%%%%%%%%%%%%%%%%%%%%%%%%%%

The simplest example of duality is between the (non)-minimal the Type-A HS gravity and ($U(N)$) $O(N)$ 
free scalars. Dropping the canonical normalization factors, the two-point functions for the 
$U(N)$ case are\footnote{In the $O(N)$ case we omit the bar over $\phi$, i.e. we have $\phi^a$, 
and the $U(N)$-invariant tensor on the r.h.s. $\delta^a_b$ is replaced by the $O(N)$-invariant 
one $\delta^{ab}$.}
\begin{align}
U(N)&: & \langle \bar{\phi}_a(\rmx) \phi^b(0)\rangle&= \delta_a^b \frac{1}{|\rmx|}\ ,
\end{align}
where $a,b=1,...,N$. As is well known \cite{Craigie:1983fb,Gelfond:2006be,Giombi:2009wh}, 
it is convenient to pack the HS currents into generating functions 
\begin{align}
j(\rmx,\eta)&= f(u,v) \bar{\phi}_a(\rmx_1) \phi^a(\rmx_2)\Big|_{\rmx_i=\rmx}\ , &&u = 
\tfrac12 \eta^\ga \eta^\gb\pl^1_{\ga\gb}\,,\quad v= \tfrac12 \eta^\ga \eta^\gb\pl^2_{\ga\gb}\ .
\end{align}
The conservation of the current implies a simple differential equation for $f$. The most convenient formulae 
are obtained \cite{Gelfond:2006be,Gelfond:2013xt} with the help of an auxiliary generating function 
$C^a(\eta|x)$ for the derivatives of the scalar field. It obeys $\pl_{\ga\gb} C^a= \tfrac{i}2 \pl_\ga\pl_\gb C^a$, 
which is solved by $C^a=\cos\left[2e^{i{\pi}/{4}}\sqrt{u}\right]\phi^a(\rmx)$.\footnote{One can put any 
factor $k$ into the definition $\pl_{\ga\gb} C^a= k \pl_\ga\pl_\gb C^a$, which 
then will alter the normalization of the correlators. Our factors are such as to facilitate the 
comparison on the both sides of AdS/CFT duality.} 
The generating function of the HS currents then has a simple factorized form:\footnote{There is another generating function that can be obtained directly from the 
conservation equation $(\pl_u +\pl_v +2 u\pl^2_u +2v \pl^2_v)f=0$, which is 
$f=e^{u-v}\cos \left(2 \sqrt{ u v}\right)$ \cite{Giombi:2009wh}. }
\begin{align}
j(\rmx,\eta)&= C_a(u)C^a(-v)=\cos[2e^{i{\pi}/{4}}\sqrt{u}]\cos[2e^{i{\pi}/{4}}\sqrt{-v}]\bar{\phi}_a(\rmx_1) \phi^a(\rmx_2)\Big|_{\rmx_i=\rmx}\ .
\end{align}
The two-point function of the currents is then\footnote{There is always an overall factor of $N$, which is present in all two-point and three-point 
functions of the HS currents. We will suppress this factor. Correlation function for specific spins $s_1,s_2,...$ is extracted out of any generating function as the coefficient of $(\eta_1)^{s_1}(\eta_2)^{s_2}...$. }
\begin{align} \label{fb2pt}
\langle j_s j_s\rangle_{f.b.}\equiv\langle j(\rmx_1,\eta_1) j(\rmx_2,\eta_2)\rangle &= 
\frac{1}{2}\frac{1}{\rmx_{12}^2}  \cosh \left(2 {P_{12}}\right)\ ,
\end{align}
where we introduced a shorthand notation $j_s$ for $j(\rmx_i,\eta_i)$ and the arguments are in 
accordance with its position inside the brackets; $f.b.$ refers to ``free boson". Sometimes we have 
to distinguish the first 
member of the family $j_0\equiv\bar{\phi}_a(\rmx)\phi^a(\rmx)$ from the others. The simplest three-point function, 
which is fixed by the symmetry, is
\begin{align}
\langle j_s j_0 j_0\rangle_{f.b.}&= \frac{2}{\rmx_{12}\rmx_{13}\rmx_{23}}\cos{\left(\tfrac12{Q_1}\right)}\ .
\end{align}
The three-point functions with two HS currents are assembled into
\begin{align}\label{bosonjjo}
\langle j_s j_s j_0\rangle_{f.b.}&= \frac{2}{\rmx_{12}\rmx_{13}\rmx_{23}}\cos{\left(\tfrac12{Q_1}
+\tfrac12 Q_2\right)}\cos P_{12}\ .
\end{align}
The three-point functions of the three HS currents are
\cite{Giombi:2010vg} (see also \cite{Giombi:2011rz,Colombo:2012jx,Didenko:2012tv,Didenko:2013bj,Gelfond:2013xt}):
\begin{align} 
%\label{threepointAnswer}
\langle j_s j_s j_s\rangle_{f.b.}&=
\frac{2}{\rmx_{12}\rmx_{23}\rmx_{31}}\cos\left(\tfrac12 Q_1+\tfrac12Q_2
+\tfrac12Q_3\right)\cos(P_{12})\cos(P_{23})\cos(P_{31})\ .
\label{3ptb}
\end{align}
It is also useful to consider the case of $U(N)$-singlet constraint with 
leftover $U(M)$ global symmetry. The generating function is
\begin{align}\label{bosonjjoU}
\langle j_s j_s j_0\rangle_{f.b.} &= \frac{2}{\rmx_{12}\rmx_{13}\rmx_{23}}
\exp\left[-\tfrac{i}{2}(Q_1+Q_2)\right]\,\cos P_{12} \ .
\end{align}
This is the most general case, all others being simple truncations. Imposing the 
bose symmetry we get \eqref{bosonjjo}. Truncation to even spins only gives the $O(N)$ case.

%%%%%%%%%%%%%%%%%%%%%%%%%%%%%%%%%%%%%%%%%%%%%%%%%%%%%%%%%%%%%%%%%%%%%%%
\subsection{Free Fermion} 
\label{subsec:freefermion}
%%%%%%%%%%%%%%%%%%%%%%%%%%%%%%%%%%%%%%%%%%%%%%%%%%%%%%%%%%%%%%%%%%%%%%%

The second example is the duality of the Type-B theory and a theory of free $U(N)$ or $USp(N)$ 
fermions, of which we consider the former. The two-point function is
\begin{align}
U(N)&: &\langle \bar{\psi}_{a\alpha}(\rmx) \psi_{\beta }^b (0)\rangle & = \delta_a^b\frac12\frac{\vec \rmx \cdot \vec \sigma_{\alpha\beta}\,\,}{|\rmx|^{3}}
=\delta_a^b\pl_{\ga\gb} |\rmx^2|^{-\tfrac12}\ .
\end{align}
HS currents are constructed analogously
\begin{align}
j(\eta,\rmx)&= f(u,v) \eta^\ga\eta^\gb \psi^a_\ga(\rmx_1)\psi_{a\gb} (\rmx_2)\Big|_{\rmx_i=\rmx}\ ,
\end{align}
and again the useful trick is to pack the derivatives into $C^a$ as follows 
\begin{align}
\pl_{\ga\gb} C^a &= \tfrac{i}2 \pl_\ga\pl_\gb C^a\ , && 
C^a=\tfrac{1}{\sqrt{u}}\sin\left[2e^{i{\pi}/{4}}\sqrt{u}\right]\psi^a_\ga(\rmx)\eta^\ga\ .
\end{align}
The generating function of the HS currents has the factorized form:
\begin{align}
j(\eta,\rmx)&= C^a(u)C_a(-v)=\frac{1}{\sqrt{-uv}}\sin\left[2e^{i{\pi}/{4}}\sqrt{u}\right]
\sin\left[2e^{i{\pi}/{4}}\sqrt{-v}\right]\psi^a_\ga(\rmx_1)\psi_{a\gb} (\rmx_2)\Big|_{\rmx_i=\rmx}\ .
\end{align}
The two-point function is normalized in the same way as that of the free boson:
\begin{align}
\langle j_s j_s\rangle_{f.f.}&= \frac{1}{2}\frac{1}{\rmx_{12}^2}  
\cosh \left(2 {P_{12}}\right)\ , && \langle\tilde{j}_0 \tilde{j}_0\rangle_{f.f.}=\frac{1}{4\rmx^4}\ ,
\end{align}
where $f.f.$ refers to ``free fermion". Here the scalar singlet operator 
\be
\tilde{j}_0=\bar{\psi}_a\psi^a 
\ee
has dimension $2$ and is not captured by the generating function above. 
The three-point function of $\tilde{j}_0$ vanishes due to parity
\begin{align}
\langle \tilde{j}_0 \tilde{j}_0 \tilde{j}_0\rangle_{f.f.}&=0\,.
\end{align}
For the other cases we find (see also \cite{Giombi:2011rz,Colombo:2012jx,Didenko:2012tv,Didenko:2013bj,Gelfond:2013xt}):
\begin{align}\label{fermionjjo}
\langle  j_s j_s\tilde{j}_0\rangle_{f.f.}&=\frac{2\cos(\tfrac12 Q_1+\tfrac12 Q_2)}
{\rmx^2_{23}\rmx_{13}^2}S_3 \sin P_{12}\ ,
\\
\langle  j_s\tilde{j}_0 \tilde{j}_0\rangle_{f.f.} & =
\frac{2\sin \tfrac12 Q_1}{\rmx_{12}\rmx^3_{23}\rmx_{31}} Q_1\ ,
\end{align}
and 
\begin{align} 
\langle j_s j_s j_s\rangle_{f.f.}&= \frac2{\rmx_{12}\rmx_{23}\rmx_{31}}\sin(\tfrac12Q_1
+\tfrac12Q_2+\tfrac12Q_3)\sin(P_{12})\sin(P_{23})\sin(P_{31})\ .
\label{3ptf}
\end{align}
Note that the expression for the correlators of $j_s$ is valid for $s\ge 1$ only, 
while any insertion of $\tilde{j}_0$ should be treated separately. Again, it is useful to work with the free fermion with leftover $U(M)$ global symmetry and the generating function is
\begin{align}
\langle  j_s j_s\tilde{j}_0\rangle_{f.f.}&=\frac{2}{\rmx^2_{23}\rmx_{13}^2}\exp(\tfrac{i}2 Q_1+\tfrac{i}2 Q_2)S_3 \sin P_{12}\ . 
\label{freefermjooU}
\end{align}

%%%%%%%%%%%%%%%%%%%%%%%%%%%%%%%%%%%%%%%%%%%%%%%%%%%%%%%%%%%%%%%%%%%%%%%
\subsection{Critical Boson} 
\label{subsec:criticalboson}
%%%%%%%%%%%%%%%%%%%%%%%%%%%%%%%%%%%%%%%%%%%%%%%%%%%%%%%%%%%%%%%%%%%%%%%

Critical Boson theory is the IR fixed point under a double-trace deformation $(\phi^2)^2$. 
The regime that is relevant for AdS/CFT is the large-$N$, see e.g. \cite{Vasiliev:1981yc,Derkachov:1997ch} 
for the systematic $1/N$ expansion. To the leading order in $1/N$ the correlation function of 
HS currents $j_s$, $s>0$ stays the same as in the free boson theory, i.e. \eqref{3ptb}. The dimension 
of the lowest singlet operator $\phi^2$, which is usually dubbed $\sigma$, jumps from $1$ for free boson to $2+\mathcal{O}(1/N)$ for critical one. 
Therefore, the spectrum of the singlet operators in the critical boson  theory looks like that 
of the free fermion in $N=\infty$ limit. For this reason the scalar singlet is denoted as $\tilde{j}_0$. Correlation 
functions with a number of $\tilde{j}_0$ insertions are related to those in the free boson theory by 
attaching propagators of the $\sigma$-field. In \cite{Giombi:2016zwa} the three-point functions 
$\langle j_{s_1} j_{s_2} \tilde{j}_0\rangle$ were fixed by employing the non-conservation equation 
in the Chern-Simons matter theories. The result is\footnote{We simplify a bit the form of the result 
by noticing that the solution to the recurrence relations is easier to obtain starting from the last 
coefficients as compared to \cite{Giombi:2016zwa}.}
\begin{align}
\langle j_{s_1} j_{s_2} \tilde{j}_0\rangle_{c.b.}& 
=f_{s_1,s_2}\frac{1}{\rmx^2_{13}\rmx^2_{23}}(Q_1)^{s_1}(Q_2)^{s_2}\sum_k A_k  
\left(\frac{P_{12}^2}{Q_1Q_2}\right)^{k}\ ,
\end{align}
where $c.b.$ refers to ``critical boson", $f_{s_1,s_2}$ is a overall spin-dependent factor and the coefficients obey $A_{-1}=0$, $A_0=1$ 
and the rest are generated via
\begin{align}
A_{n}&={\tiny\frac{A_{n-1} \left(s_1s_2+(2n-1-s_1)(2n-1-s_2)-5n+4)\right)
-2 A_{n-2} (n-s_1-2) (n-s_2-2)}{n (2 n-1)} }\ . \notag
\end{align}
The coefficients $f_{s_1,s_2}$ depend on normalization of $\langle j_s j_s\rangle_{c.b.}$ two-point 
functions. We choose the same normalization as in the free boson theory \eqref{fb2pt}, which gives:
\begin{align}
f_{s_1,s_2}&=-\frac{i^{s_1+s_2} 2^{-s_1-s_2+2}}{\Gamma \left(s_1+\frac{1}{2}\right) 
\Gamma \left(s_2+\frac{1}{2}\right)}\ .
\end{align}

%%%%%%%%%%%%%%%%%%%%%%%%%%%%%%%%%%%%%%%%%%%%%%%%%%%%%%%%%%%%%%%%%%%%%%%
\subsection{Critical Fermion}
\label{subsec:criticalfermion}
%%%%%%%%%%%%%%%%%%%%%%%%%%%%%%%%%%%%%%%%%%%%%%%%%%%%%%%%%%%%%%%%%%%%%%%

Critical fermion is the Gross-Neveu model, i.e. free fermion with $(\bar{\psi}\psi)^2$-deformation 
added. The large-$N$ expansion works fine despite the apparent non-renormalizability of the 
interactions, see e.g. \cite{Muta:1976js,Manashov:2016uam}. Again, the correlation functions of 
HS currents $j_s$, $s>0$ are the same as in the free fermion theory \eqref{3ptf} to the leading 
order in $1/N$, while the dimension of $\sigma=\bar{\psi}\psi$ jumps from $2$ in the free theory to $1+\mathcal{O}(1/N)$ 
in the Gross-Neveu one. Therefore, the spectrum of singlet operators looks like that of the free 
boson theory in the $N=\infty$ limit and we use the same notation ${j}_0$ for $\bar{\psi}\psi$. The three-point functions 
$\langle j_{s_1} j_{s_2} {j}_0\rangle$ were found in \cite{Giombi:2016zwa} to have the form:
\begin{align}
\langle j_{s_1} j_{s_2} {j}_0\rangle_{c.f.}&=g_{s_1,s_2}\frac{1}{\rmx_{13}\rmx_{23}\rmx_{12}}S_3 P_{12} (Q_1)^{s_1-1}(Q_2)^{s_2-1}\sum_k A_k  \left(\frac{P_{12}^2}{Q_1Q_2}\right)^{k}\ ,
\end{align} 
where $c.f.$ refers to ``critical fermion", $A_{-1}=0$, $A_0=1$ and the rest of $A_k$ is generated via
\begin{align}
A_{n}&=-\frac{8 A_{n-2} (-n+s_1+1) (-n+s_2+1)-4 A_{n-1} ((2 n-s_1-1) (2 n-s_2-1)+n+s_1 s_2)}{4 n (2 n+1)}\ . \notag
\end{align}
The overall factor in our normalization is
\begin{align}
g_{s_1,s_2}&=-\frac{i^{s_1+s_2} 2^{-s_1-s_2+2}}{\Gamma \left(s_1+\frac{1}{2}\right) 
\Gamma \left(s_2+\frac{1}{2}\right)}\ .
\end{align}

%%%%%%%%%%%%%%%%%%%%%%%%%%%%%%%%%%%%%%%%%%%%%%%%%%%%%%%%%%%%%%%%%%%%%%%
\subsection{Chern-Simons Matter Theories} 
\label{subsec:csmatter}
%%%%%%%%%%%%%%%%%%%%%%%%%%%%%%%%%%%%%%%%%%%%%%%%%%%%%%%%%%%%%%%%%%%%%%%

There are four theories that are obtained by coupling the free/critical boson/fermion theories 
discussed above to Chern-Simons gauge field. The conjecture of three-dimensional bosonization is that they are equivalent in pairs under an appropriate identification of parameters 
$N$ and $\lambda=N/k$. We are interested in $\langle j_{s_1} j_{s_2} {j}_0\rangle$ and 
$\langle j_{s_1} j_{s_2} \tilde{j}_0\rangle$ correlators, which are constrained 
to be \cite{Maldacena:2012sf}:\footnote{There is some ambiguity in the overall factor due to the choice of normalization of the scalar singlet operator.}
\besubeqs\label{jjo}
\begin{align}
\langle j_{s_1} j_{s_2} j_0\rangle_{B.}&= {\tilde{N}} 
\left[\cos\theta \langle j_{s_1} j_{s_2} {j}_0\rangle_{f.b.}
+\sin \theta\langle j_{s_1} j_{s_2} j_0\rangle_{odd}\right]\ ,
\w2
\langle j_{s_1} j_{s_2} \tilde{j}_0\rangle_{F.}&= 
{ \tilde{N}} \left[\cos\theta \langle j_{s_1} j_{s_2} \tilde{j}_0\rangle_{f.f.}
+\sin \theta\langle j_{s_1} j_{s_2} \tilde{j}_0\rangle_{odd}\right]\ ,
\end{align}
\esubeqs
where $B.$ and $F.$ refer to CS-boson and CS-fermion, respectively. Here $\tilde{N}$ and $\theta$ are two macroscopical parameters and we recall that $\cos^2\theta=1/(1+\tilde{\lambda}^2)$. This result holds true to the leading order in $\tilde{N}$ but to all orders in $\tilde{\lambda}$. The expressions in terms 
of the microscopical parameters depend on type of theory and on $N$, $\lambda=N/k$. We can 
work with $\tilde{N}$ and $\theta$ since their microscopical origin is invisible from the bulk. 
Note that the odd structures are different in the two cases (they even have different 
conformal dimensions due to $j_0$ and $\tilde{j}_0$). In fact the two odd structures can 
be found as the regular correlators on the opposite side of the bosonization duality:
\begin{align}
\langle j_{s_1} j_{s_2} j_0\rangle_{odd}&=\langle j_{s_1} j_{s_2} {j}_0\rangle_{c.f.} \,,&
\langle j_{s_1} j_{s_2} \tilde{j}_0\rangle_{odd}&=\langle j_{s_1} j_{s_2} \tilde{j}_0\rangle_{c.b.}\ .
\label{parityodd}
\end{align}
As far as $\langle j_{s_1} j_{s_2} \tilde{j}_0\rangle$ correlators are concerned, all the 
results reviewed in the previous sections follow from \eqref{jjo} and this is the structure we 
would like to reproduce from the bulk side, including the details of the correlators and 
normalization factors. Note that the term `parity odd correlator' corresponding to the second part 
of \eqref{jjo} may not have anything to do with parity, in view of \eqref{parityodd}. For example, 
the usual parity even correlator $\langle j_s j_s \tilde{j}_0\rangle_{c.b.}$ in the critical 
boson theory appears to be odd from the point of view of the dual fermionic theory.

%%%%%%%%%%%%%%%%%%%%%%%%%%%%%%%%%%%%%%%%%%%%%%%%%%%%%%%%%%%%%%%%%%%%%%%
\section{Higher Spin Interactions} 
\label{sec:hsinteractions}
%%%%%%%%%%%%%%%%%%%%%%%%%%%%%%%%%%%%%%%%%%%%%%%%%%%%%%%%%%%%%%%%%%%%%%%

We briefly review the structure of the equations that results from the Vasiliev equations 
\cite{Vasiliev:1990en}. For more detailed reviews, we refer to 
\cite{Vasiliev:1999ba,Didenko:2014dwa,Giombi:2016ejx}. The corrections to the free 
equations that are bilinear in the fields were worked out in 
\cite{Sezgin:2002ru,Giombi:2009wh,Didenko:2014dwa,Boulanger:2015ova}. 
The main conclusion that is based on \cite{Giombi:2009wh,Boulanger:2015ova} is that, up to quadratic order, 
the Klein-Gordon equation is sourced by two type of terms. One part is fixed by the HS algebra and is local 
enough for the computation of the correlation functions using field theory tools. A second part gives rise to infinities,
though  a proposal has been made \cite{Vasiliev:2016xui} on how to obtain finite results by a set of field redefinitions. 
While our computations will mostly be based on the use of the first part, we shall nonetheless test 
this proposal as well in Section \ref{sec:vertices}, and comment further about it in Section \ref{sec:discussion}.

%%%%%%%%%%%%%%%%%%%%%%%%%%%%%%%%%%%%%%%%%%%%%%%%%%%%%%%%%%%%%%%%%%%%%%%
%\subsection{General Structure of the Unfolded Equations} 
%%%%%%%%%%%%%%%%%%%%%%%%%%%%%%%%%%%%%%%%%%%%%%%%%%%%%%%%%%%%%%%%%%%%%%%

The convenient field variables that a HS theory can be built with are the Fronsdal fields. 
The Vasiliev equations yield first order differential equations, which upon solving for the 
auxiliary fields give equations in term of the Fronsdal fields and an infinitely many of 
differential consequences of these equations, which we drop. 

For the purpose of computing tree-level Witten diagrams it is sufficient to impose the transverse 
and traceless gauge. Then the $4d$ free Fronsdal equations for spins $s\ge 1$ read:\footnote{Our normalization of 
the cosmological constant differs by a factor of two from the canonical one, which is the price 
to pay for the most natural coefficients in the $sp(4)$-flatness condition that provides us 
with the vierbein and spin-connection of $AdS_4$. See Appendix A for further details.}
\begin{align}
(\square +2(s^2-2s-2))\Fron_{\mm_1...\mm_s}&=0\ , && \Fron\fud{\nn}{\nn\mm_3...\mm_s}=0\ , 
&& \nabla^\nn\Fron_{\nn\mm_2...\mm_s}=0\ .
\end{align}
To make a link to the frame-like and then to the unfolded formulation of the $4d$ HS theories 
we replace a set of the traceless world tensors $\Fron_{\mm_1...\mm_s}$ with the generating function 
in the spinorial language as\footnote{We note that the Fronsdal field is now a fiber (spin)-tensor, which is the world 
one contracted with $s$ vierbeins. }
\begin{align}
\label{spinorialFrfield}
\Fron (y,\bry|x) &=\sum_{s} \frac{1}{s!s!}\Fron_{\ga_1...\ga_s,\gad_1...\gad_s} (x) \, y^{\ga_1}...y^{\ga_s}\,\bry^{\gad_1}...\bry^{\gad_s}\ .
\end{align}
Then the gauge-fixed Fronsdal equations, as recovered from the $4d$ HS theory, read:
\be
\left(\square  +2(N^2-2N-2) \right)\Fron = 0\ , \qquad  (\pl\nabla\bar{\pl})\Fron =0\ ,
\ee
where $N$ is the number operator $N=y^\alpha\pl_\alpha$ that counts spin (equivalently we 
can use $\bar{N}=\bry^\gad\pl_\gad$). The last equation manifests the transverse gauge. 

In the unfolded approach \cite{Vasiliev:1988sa}, a specific multiplet of HS fields is packed 
into generating functions $\omega$ and $C$ that take values in a HS algebra. 
In our case the relevant HS algebra \cite{Vasiliev:1986qx} is the even part of the Weyl 
algebra $A_2$. It is convenient to split the four generators of $A_2$ into (anti)-fundamentals 
of $sl(2,\mathbb{C})$
\begin{align}
[\hat y_\ga,\hat y_\gb]&=2i\epsilon_{\ga\gb}\,, &
[\hat \bry_\gad,\hat \bry_\gbd]&=2i\epsilon_{\gad\gbd}\,.
\end{align}
In practice, non-commuting operators $\hat y_\ga$ and $\hat\bry_\gad$ are replaced by commuting 
variables $y_\ga$, $\bry_\gad$ while the product is replaced with the star product
\begin{align}
(f\star g)(y,\bry) &= f(y,\bry)\exp i\left[\frac{\overleftarrow{\pl}}{\pl y^\ga}
\epsilon^{\ga\gb}\frac{\overrightarrow{\pl}}{\pl y^\gb}+\frac{\overleftarrow{\pl}}{\pl \bry^\gad}\epsilon^{\gad\gbd}\frac{\overrightarrow{\pl}}{\pl \bry^\gbd}\right]g(y,\bry)\ .
\end{align}
The bosonic higher spin algebra is defined as the even subalgebra, i.e. $f(y,\bry) \in hs$ 
implies $f(y,\bry)=f(-y,-\bry)$. It is straightforward to tensor any higher spin algebra 
with matrix algebra as to get $U(N)$ extension, for example. The SUSY case are studied in \cite{Konstein:1989ij}.

The field content consists of one-form $\omega=\omega_\mm(y,\bry|x)\,dx^\mm$ and zero-form 
$C=C(y,\bry|x)$. 
In free theory the Fronsdal fields can be identified with certain components of $\omega$, which 
in the traceless gauge is 
\begin{align}
h^{\ga\gad}\pl_\ga \pl_\gad \Phi (y,\bry|x) \in \omega(y,\bry|x)
\end{align}
$\omega$ and $C$ obey the following linearized equations, known as the on mass-shell theorem (OMST) \cite{Vasiliev:1988sa},
\begin{align}
\adD \omega=\mathcal{V}(\Omega,\Omega,C)\ ,\qquad \tadD C=0\ ,
\label{xpaceseqQBA}
\end{align}
where $\mathcal{V}(\Omega,\Omega,C)$ is a star function quadratic in $\Omega$ and linear in $C$, and
\begin{align}
\adD\omega &\equiv d\omega -\Omega\star \omega - \omega\star \Omega = 
\nabla\omega -h^{\ga\gad}(y_\ga\bar\pl_\gad+\bry_\gad\pl_\ga)\, \omega\ ,
\label{AdjointDer}
\w2
\tadD C &\equiv dC -\Omega\star C +  C \star \pi(\Omega) 
=\nabla C + ih^{\ga\gad}(y_\ga\bry_\gad-\pl_\ga\bar\pl_\gad)\, C\ .
\label{TwistedDer}
\end{align}
The $AdS_4$ connection $\Omega$ is defined in  Appendix A, and the Lorentz covariant 
derivative acts in the same way on $\omega(y,\bar y|x)$ and $C(y,\bar y|x)$ as
\be
\nabla  \equiv d -\omega_{(0)}^{\ga\gb} y_\ga \pl_\gb
-{\bar\omega}_{(0)}^{\gad\gbd}\bry_\gad \bar \pl_\gbd\ .
\label{LorentzDer}
\ee
The difference between $\adD$ and $\tadD$ is due to automorphism $\pi$: $\pi(f)(y,\bry)=f(y,-\bry)=f(-y,\bry)$.
The vertex that relates the order-$s$ curl of the Fronsdal field 
to the HS Weyl tensors reads
\begin{align}
\label{OMSTcocycle}
\mathcal{V}(\Omega,\Omega,C)&=A\big[ H^{\ga\gb}\pl_\ga\pl_\gb C(y,0)e^{-i\theta}+
 \bar H^{\gad\gbd}\bar\pl_\gad\bar\pl_\gbd C(0,\bry)e^{i\theta}\big]\ ,
\end{align}
where $H^{\ga\gb}=h\fud{\ga}{\gamma}\wedge h^{\gb\gamma}$, and 
analogously for $\bar H^{\gad\gbd}$. The constant $A$ is an arbitrary normalization factor that we choose to be {\footnotesize{$A=i/4$}}. 
The linearized equations \eq{xpaceseqQBA} are invariant under the linearized HS 
gauge transformations
\be
\delta \omega=\adD\xi\ ,  \qquad \delta C=0\ .
\ee
Equations \eqref{xpaceseqQBA} are equivalent to Fronsdal equations supplemented with differential consequences thereof. Up to the second order the unfolded equations should have the following schematic form
\besubeqs
\begin{align}
d\omega&=\omega\star \omega + \mathcal{V}(\omega,\omega,C)+\mathcal{V}^2(\omega,\omega,C,C)
+{\cal O}(C^3)\,,
\\
dC&=\omega\star C-C\star \pi(\omega)+\mathcal{U}(\omega,C,C)+ {\cal O} (C^3)\,,\label{UCC}
\end{align}
\esubeqs
where the vertices $\mathcal{V}^2$ and $\mathcal{U}$ need to be specified. The free equations result upon substituting $\omega\rightarrow \Omega+\omega$ and picking the terms that are linear in $\omega$ and $C$:
\begin{align}
&\adD \omegaone=\mathcal{V}(\Omega,\Omega,\Cone)\ , &
&\tadD \Cone=0\ .
\end{align}
At the second order the weak-field expansion over the AdS background leads to
\besubeqs
\begin{align}
&\adD \omegatwo-\mathcal{V}(\Omega,\Omega,\Ctwo)=\omegaone\star\omegaone+
2\mathcal{V}(\Omega,\omegaone,\Cone)+ \mathcal{V}(\Omega,\Omega,\Cone,\Cone)
\label{xpaceseqBA}\ ,
\\
&\tadD \Ctwo=\omegaone\star \Cone-\Cone\star \pi(\omegaone)+\mathcal{U}(\Omega,\Cone,\Cone)
\label{xpaceseqBB}\ .
\end{align}
\esubeqs
We see that the second order fluctuations are sourced by the terms 
that are bilinear in the first order fluctuations. The terms that are bilinear in the 
zero-forms, the $CC$-terms for short, can be non-local. 
The part of the equations that does not have any problems with locality comes from 
the commutator in the HS algebra
\begin{align}\label{maineq}
&\tadD \Ctwo=\omegaone\star \Cone-\Cone\star \pi(\omegaone) + {\cal O}(C^2)\ .
\end{align}
For some values of spins we do not expect the $CC$-terms to contribute. Indeed, 
when one of the legs is scalar, there is a unique (parity-preserving) coupling $s_1-s_2-0$, \cite{Metsaev:2005ar}. 
It is non-abelian whenever $s_1\neq s_2$ and is abelian for $s_1=s_2$.

Let us note that the computation of correlators from the equations of motion that are not 
derived from an action principle does not guarantee the bose symmetry of the correlators 
under the permutation of the points. The same correlator $\langle j_{s_1} j_{s_2} j_0\rangle$ 
can be obtained in two different ways: either by treating $j_{s_1}$, $j_{s_2}$ as sources and 
then solving for the scalar field or by treating $j_{s_1}$, $j_0$ as sources and solving for 
the spin-$s_2$ Weyl tensor. Both computations are possible with \eqref{maineq}. It was checked 
in \cite{Giombi:2009wh} for the leading coefficients that the two ways of getting the same 
correlator give a bose symmetric correlator. 

Lastly, Eq.\eqref{maineq} remains consistent if the fields $\omega$, $C$ are extended to matrix-valued fields. This way one can introduce Yang-Mills groups on top of the HS algebra. On the CFT side this should correspond to leftover global symmetries, i.e. those symmetries that remain after the singlet constraint is imposed.

%%%%%%%%%%%%%%%%%%%%%%%%%%%%%%%%%%%%%%%%%%%%%%%%%%%%%%%%%%%%%%%%%%%%%%%
\section{Kinematics of the Boundary-to-Bulk Propagators} 
\label{sec:Kinematic}
%%%%%%%%%%%%%%%%%%%%%%%%%%%%%%%%%%%%%%%%%%%%%%%%%%%%%%%%%%%%%%%%%%%%%%%

In this section we discuss the boundary-to-bulk propagators for HS  fields. The problem has 
been extensively studied starting from the lower spin fields, see e.g. \cite{DHoker:2002nbb}. 
Specifically, we need the unfolded propagators, i.e. the propagators for the Fronsdal fields 
supplemented with derivatives thereof as to obey the free equations \eqref{xpaceseqQBA}. 
In some form the unfolded propagators were found in \cite{Giombi:2009wh}. In \cite{Didenko:2012tv} 
it was observed that the propagators are simple functions that depend on a few universal 
geometrical data, which is the spinorial analog of \cite{Allen:1985wd}. 

%%%%%%%%%%%%%%%%%%%%%%%%%%%%%%%%%%%%%%%%%%%%%%%%%%%%%%%%%%%%%%%%%%%%%%%
\subsection{Definitions} 
%%%%%%%%%%%%%%%%%%%%%%%%%%%%%%%%%%%%%%%%%%%%%%%%%%%%%%%%%%%%%%%%%%%%%%%

One of the basic objects is the Witten bulk-to-boundary propagator for the scalar field 
from the bulk point $x^\mm=(\rmx^i,z)$ to the boundary point $\rmx_a$
%
% \footnote{ $\rmx^i \rmx^j\eta_{ij}=-\frac12 \rmx_{\ga\gb}\rmx^{\ga\gb}$, where $\eta$ is 
% the Lorentz metric on the boundary.} [Already given in Appendix A.]
%
\begin{align}
K_a&=\frac{z}{(\rmx-\rmx_a)^2+z^2}\,. %\qquad K_0 \equiv K\ ,
\end{align}
We often set $\rmx_a=0$ in this section as the generic point can be recovered thanks 
to the three-dimensional Poincar\'e invariance. The propagator is the boundary limit of 
the geodesic distance, which the bulk-to-bulk propagator should depend on. It obeys the 
regular ($\Delta=1$) boundary condition. The propagator $K$ can be used to define a wave-vector, 
which turns into bi-spinor $F^{\ga\gad}$ in the $4d$ spinorial language:
\begin{align}
d \ln{K}&=F_{\ga\gad} h^{\ga\gad}\ ,
\label{AppbtobB}
\end{align}
where
\begin{align}
F^{\ga\gad}&=\left(\frac{2z}{\rmx^2+z^2}\rmx^{\ga\gad}-
\frac{\rmx^2-z^2}{\rmx^2+z^2}\,i\epsilon^{\ga\gad}\right)\ .
\end{align}
It will play the same role as the on-shell momentum $p$ plays for $e^{ipx}$ and is 
the boundary limit of the vector that is tangent to the geodesic connecting two bulk 
points. There also exist the parallel-transport bi-spinors $\Pi^{\ga\gb}$ and 
$\Pib^{\gad\gb}$ given by
\begin{align}\label{Xidef}
\Pi^{\ga\gb}&=K \left(\frac{1}{\sqrt{z}}\,
\rmx^{\ga\gb}+\sqrt{z}\,i\epsilon^{\ga\gb}\right)\ , &
\Pib^{\gad\gb}&=K \left(\frac{1}{\sqrt{z}}\,
\rmx^{\gad\gb}-\sqrt{z}\,i\epsilon^{\gad\gb}\right)=(\Pi^{\ga\gb})^\dag\ ,
\end{align}
that allow one to propagate the boundary polarization spinors into the bulk:
\begin{align}
%\label{Xidef}
\xi^\ga &= \Pi^{\ga\gb}\eta_\gb\,e^{+i\frac{\pi}4} \ , \qquad 
\bar{\xi}^{\gad}=(\xi^\ga)^\dag=\Pib^{\gad\gb}\eta_\gb\,e^{-i\frac{\pi}4} \ .
\end{align}
This is the full set of the data that any propagator can depend on. The set is closed 
under covariant derivatives:
\begin{align}
\nabla F^{\ga\gad}=h^{\ga\gad}+F\fud{\ga}{\gdd}h^{\gd\gdd}F\fdu{\gd}{\gad}=0\ ,
\end{align}
and the parallel transported spinors obey
\begin{align}
\nabla\xi^\ga-F\fud{\ga}{\gdd} \xi_\gd h^{\gd\gdd}=0\ ,\qquad
\nabla\xi^{\gad}-F\fdu{\gd}{\gad} \xi_{\gdd} h^{\gd\gdd}=0\ .
\label{AppbtobD}
\end{align}
In practice it is useful to rewrite the Lorentz-covariant derivatives with all 
indices being explicit:
\begin{align}
\nabla_{\ga\gad}K &= K F_{\ga\gad}\ ,& \nabla_{\ga\gad} \xi_\beta &=F_{\beta\gad}\xi_\ga\  , &
\nabla_{\ga\gad} \brxi_{\dot\beta} &=F_{\ga\dot\beta}\xi_\gad\ , &
\nabla_{\ga\gad} F_{\gb\gbd} &=2\epsilon_{\ga\gb}\epsilon_{\gad\gbd}+F_{\ga\gad}F_{\gb\gbd}\ .
\end{align}
As a consequence of the differential constraints above one also finds
\begin{align}
(\square-4) K &=0\ , & (\square-6)F^{\ga\gad} &=0\ ,
& (\square-4)\xi^{\ga}&=0\ , & (\square-4)\brxi^{\gad}&=0\ .
\end{align}
%
% Let us note that the bulk-to-boundary objects introduced above can be used for all fields in $AdS_4$, 
% massless or massive. 

%%%%%%%%%%%%%%%%%%%%%%%%%%%%%%%%%%%%%%%%%%%%%%%%%%%%%%%%%%%%%%%%%%%%%%%
\subsection{Algebraic Identities} 
%%%%%%%%%%%%%%%%%%%%%%%%%%%%%%%%%%%%%%%%%%%%%%%%%%%%%%%%%%%%%%%%%%%%%%%

The quantities defined above obey several algebraic identities. The wave-vector satisfies 
\be
F^{\ga\gad}F\fud{\gb}{\gad}=\epsilon^{\ga\gb}\ ,\qquad
F^{\ga\gad}F\fdu{\ga}{\gbd}=\epsilon^{\gad\gbd}\ .
\label{FisSpTwo}
\ee
It behaves in many respects as a 'symplectic' structure that converts
the dotted and undotted indices to each other as follows
\begin{align}
F\fud{\ga}{\gdd}\Pib^{\gdd\gb} &= i\Pi^{\ga\gb}\ , 
& F\fdu{\gc}{\gad}\Pi^{\gc\gb} &=-i\Pib^{\gad\gb}\ , 
& \Pi\fud{\ga}{\gc}\Pib^{\gad\gc}=-iKF^{\ga\gad}\ ,
& \\
\xi^{\ga} &=F\fud{\ga}{\gad}\bar{\xi}^\gad\ , 
& \bar{\xi}^{\gad} &={\xi}^\ga F\fdu{\ga}{\gad}\ .
\end{align}
The parallel-transport bi-spinors $\Pi^{\ga\gb}$ and $\Pib^{\gad\gb}$ obey identities 
similar to \eqref{FisSpTwo}:
\begin{align}
\Pi^{\ga\gb}\Pi\fud{\gc}{\gb} &=K \epsilon^{\ga\gc}\ , &
\Pib^{\gad\gb}\Pib\fud{\dot\gc}{\gb} &=K \epsilon^{\gad\dot\gc}\ ,
\\
\Pi^{\gb\ga}\Pi\fdu{\gb}{\gc} &=K \epsilon^{\ga\gc}\ , &
\Pib^{\gbd\ga}\Pib\fdu{\gbd}{\gc} &=K \epsilon^{\ga\gc}\ .
\end{align}
There are also useful identities involving the one-form 
$h^{\alpha\dot\alpha} =dx^\mm h_\mm^{\alpha\dot\alpha}$:
\begin{align}
&(F\cdot h) F^{\ga\gad}+h^{\ga\gad}=F\fud{\ga}{\gbd}h^{\gb\gbd}F\fdu{\gb}{\gad}\ ,
\\
&(F\cdot h) \xi^\alpha +(F\fud{\ga}{\gdd}h\fdu{\gb}{\gdd}-F_{\gb\gdd}h^{\ga\gdd})\xi^\gb=0\ ,
\end{align}
which result from the fact that anti-symmetrization over any three spinorial indices 
vanishes identically.

%%%%%%%%%%%%%%%%%%%%%%%%%%%%%%%%%%%%%%%%%%%%%%%%%%%%%%%%%%%%%%%%%%%%%%%
\subsection{Inversion Map}
\label{inv}
%%%%%%%%%%%%%%%%%%%%%%%%%%%%%%%%%%%%%%%%%%%%%%%%%%%%%%%%%%%%%%%%%%%%%%%

The basic computational tool we will employ is based on the inversion trick \cite{Freedman:1998tz}. 
For that reason it is important to know the transformation properties of the variables defined 
above under the inversion isometry of $AdS_4$. Since the boundary-to-bulk objects 
$K$, $F$, $\xi$ and $\bar{\xi}$ depend both on the bulk point, on the boundary point 
and polarization spinor $\eta$, the inversion map on the $AdS$-side should be accompanied 
by the inversion map on the boundary. Using the inversion map rules \eq{prime},
we derive the following transformation properties:
\besubeqs
\begin{align}
K(R(\rmx,z);R\rmx_i)&= \rmx_i^2 K(\rmx,z;\rmx_i)\ ,
\w2
\Pi^{\ga\beta}(R(\rmx,z);R(\rmx_i))&
=-J\fud{\ga}{\gdd}\bar{\Pi}\fud{\gdd}{\delta}(\rmx,z;\rmx_i)\rmx_i^{\delta\beta}\ ,
\w2
\xi^{\ga}(R(\rmx,z);R(\rmx_i,\eta_i))&=+iJ\fud{\ga}{\gdd}\bar{\xi}^{\gdd}(\rmx,z;\rmx_i,\eta_i)\ ,
\w2
\bar{\xi}^{\gad}(R(\rmx,z);R(\rmx_i,\eta_i))&=-iJ\fdu{\gc}{\gad}{\xi}^{\gc}(\rmx,z;\rmx_i,\eta_i)\ ,
\w2
F^{\ga\gad}(R(\rmx,z);R(\rmx_i))&=J\fud{\ga}{\gbd} J\fdu{\gb}{\gad}F^{\gb\gbd}(\rmx,z;\rmx_i)\ ,
\end{align}
\esubeqs
where we defined $J^{\ga\gad}$ as
\begin{align}
J^{\ga\gad}&=\frac{x^{\ga\gad}}{\sqrt{x^2}}=\frac{\rmx^{\ga\gad}
+iz\epsilon^{\ga\gad}}{\sqrt{\rmx^2+z^2}}\,, && J\fud{\ga}{\gdd}J^{\gb\gdd}=-\epsilon^{\ga\gb}\ .
\end{align}
%

%%%%%%%%%%%%%%%%%%%%%%%%%%%%%%%%%%%%%%%%%%%%%%%%%%%%%%%%%%%%%%%%%%%%%%%
\section{Vertices and Propagators}
\label{sec:vertices}
%%%%%%%%%%%%%%%%%%%%%%%%%%%%%%%%%%%%%%%%%%%%%%%%%%%%%%%%%%%%%%%%%%%%%%%

In this section we discuss the boundary-to-bulk propagators for HS fields and evaluate the 
vertex \eqref{maineq} on the propagators. 

%%%%%%%%%%%%%%%%%%%%%%%%%%%%%%%%%%%%%%%%%%%%%%%%%%%%%%%%%%%%%%%%%%%%%%%
\subsection{Propagators}
%%%%%%%%%%%%%%%%%%%%%%%%%%%%%%%%%%%%%%%%%%%%%%%%%%%%%%%%%%%%%%%%%%%%%%%
With the help of the geometric objects introduced in Section \ref{sec:Kinematic} it is very 
easy to construct propagators. First of all, there is a unique expression for the Fronsdal 
field propagator:
\begin{align}
\Fron_{\ga(s),\gad(s)}&=-2\sigma^{2s} A \frac{\Gamma[s]\Gamma[s]}{\Gamma[2s]} 
\left[ K \xi_{\ga(s)} \brxi_{\gad(s)} \right]\ ,
\end{align}
where $A$ is the factor from the free unfolded equations \eqref{OMSTcocycle}, $\sigma$ is a parameter that 
counts spin, the $\Gamma$-functions will be explained later as the most convenient normalization.

In practice we need the propagators for $\omega$ and $C$ fields that enter the unfolded equations. 
These fields encode derivatives of the Fronsdal field.  The propagators can be written in a very 
compact form as\footnote{The unfolded propagators were first found in \cite{Giombi:2009wh} in a 
different form, especially the $\omega$ propagator where the gauge ambiguity is essential. 
The $C$ propagator was cast into the form below in \cite{Didenko:2012tv}, see also 
\cite{Giombi:2010vg,Didenko:2013bj,Colombo:2012jx}. The simple expression for the $\omega$ 
propagator was obtained together with Slava Didenko and was not published by us before.} 
\begin{align}
\omega&=-2\sigma^2 AK h_{\ga\gad}\xi^\ga\brxi^\gad\int_0^1 dt\nonumber\, 
\exp{i[\sigma ty^\ga \xi_\ga-(1-t)\sigma \bry^\gad \brxi_\gad]}\ ,
\\
C &= K \exp{i [-y_\ga F^{\ga\gad} \bry_\gad+\sigma y^\ga \xi_\ga
+\theta]}+K \exp{i[- y_\ga F^{\ga\gad} \bry_\gad-\sigma\bry^\gad \brxi_\ga-\theta]}\ ,
\\
&=K \exp{i [-y_\ga F^{\ga\gad} \bry_\gad+\sigma y^\ga \xi_\ga+\theta]}+h.c.\ ,\nonumber
\end{align}
where $A$ is a constant that is related to the normalization of the free unfolded 
equations \eqref{OMSTcocycle}. The $h.c.$ operation is defined as
\begin{align}
h.c.(\xi)&=-\brxi\,, && h.c. (\theta)=-\theta\,
\end{align}
and $\sigma$ is just a factor that counts spins, so can be put 
to one or any $\sigma(s)$. We will fix this normalization later.

%%%%%%%%%%%%%%%%%%%%%%%%%%%%%%%%%%%%%%%%%%%%%%%%%%%%%%%%%%%%%%%%%%%%%%%
\subsection{Vertices}
\label{subsec:vertices}
%%%%%%%%%%%%%%%%%%%%%%%%%%%%%%%%%%%%%%%%%%%%%%%%%%%%%%%%%%%%%%%%%%%%%%%

Let us remind that the equation we will extract the correlation function from is 
\begin{align}
\tadD \Ctwo&=\omega\star C-C\star \pi(\omega)+ {\cal O}(C^2)\ ,
\label{Ctwoeq}
\end{align}
where we evaluate the r.h.s. on the propagators and then solve for the Klein-Gordon 
equation. Expanding the master zero-form $C$ as 
\be
C(y,\bar y|x) = \phi(x) + \phi_{\ga\gad}(x) y^\ga\bry^{\gad}+\cdots\ ,
\ee
the Klein-Gordon equation with a source is hidden in the first order form:
\begin{align}
d\phi-i h_{\ga\gad} \phi^{\ga\gad}&=h^{\ga\gad}P_{\ga\gad}\ ,
\\
\nabla \phi^{\ga\gad}+h^{\ga\gad} \phi+...&=h^{\gb\gbd} P^{\ga\gad}_{\gb\gbd}\ ,
\end{align}
where the one-form $\mathcal P =h^{\ga\gad}P_{\ga\gad}$ denotes the source built out of the free fields:
\begin{align}
\mathcal P &=\omega\star C-C\star \pi(\omega)
\end{align}
and we need the first two Taylor coefficients only:
\be
\mathcal P=  h^{\ga\gad} P_{\ga\gad}(y,\bry) = h^{\ga\gad}\left[P_{\ga\gad} + P^{\gb\gbd}_{\ga\gad}\,y_\gb\bry_{\gbd}+\cdots\right]\ .
\ee
The first constraint can be solved for the auxiliary field $\phi^{\ga\gad}$ and the result 
then plugged into the second one to get the Klein-Gordon equation with a source \cite{Sezgin:2003pt}:
\besubeqs\label{vertexAA}
\begin{align}
(\square -4)\phi&=\Big[\nabla^{\ga\gad}P_{\ga\gad}+i \pl^\ga\pl^\gad P_{\ga\gad}\Big]\Big|_{y,\bry=0}\ .
\end{align}
Since the propagators are known, we can get the source explicitly as
\begin{align}
(\square -4)\phi &=-4 A K_1K_2 \int_0^1 dt\,\big[(1+h.c.)
(1+\pi_{\xi_1})\left(w+i t(1-t)w^2+itw\xi_1\xi_2 \right)B\big]\label{vertexraw}\ ,
\\
B &=\exp i[t(1-t) w +t \xi_1\xi_2+\theta]\ ,
\\
w &= (\xi_1 F_2 \brxi_1)\ ,
\end{align}
\esubeqs
where $K_1$, $\xi_1$ refer to $\omega$ and $K_2$, $\xi_2$, $F_2$ to $C$, and 
\be
(\xi_1\xi_2) \equiv \xi_1^\alpha \xi_{2\alpha}\ , \qquad  (\xi_1 F_2 \bar{\xi}_1)
\equiv \xi_{1\ga} F^{\ga\gad} \xi_{1\gad}\ .
\ee
Here $\pi_\xi$ is the twist map that is now realized on $\xi$ (or $\brxi$ due to 
the bosonic projection), $\pi(\omega)=\omega(-\xi,\brxi)=\omega(\xi,-\brxi)$. 
The appearance of $\pi_{\xi_1}$ is responsible for the 
difference between HS fields with and without additional Yang-Mills groups. 
If we keep $\pi_{\xi_1}$ then in the dual CFT the correlators 
$\langle j_{s_1}j_{s_2} j_0\rangle$ will vanish for $s_1+s_2$ odd. If we drop 
$\pi_{\xi_1}$ then on the CFT side $\langle j_{s_1}j_{s_2} j_0\rangle$ does not 
vanish for $s_1+s_2$ odd and therefore we have a leftover global symmetry group. 
The latter case is more general and is easier to deal with. 

On expanding the generating function \eqref{vertexraw} and picking the terms of spins 
$s_1$ from $\omega$ and $s_2$ from $C$ we find the vertex evaluated on the propagators 
in a very simple form\footnote{This vertex is also present in \cite{Giombi:2009wh}, but 
it does not seem to have such a simple form as below, which should be related to the 
$\omega$ propagator being in a different gauge.}
\besubeqs\label{vertexmain}
\begin{align}
(\square -4)\phi &=-4 A K_1K_2 \sum_{s_1,s_2} V_{s_1,s_2,0}\ ,
\label{vertex1}
\\
V_{s_1,s_2,0}&=  v_{s_1,s_2,0}\big[ (\xi_1 F_2 \brxi_1)^{s_1-s_2} 
(\xi_1\xi_2)^{2s_2} e^{i\theta}+h.c.\big]\ ,
\label{vertex2}
\\
v_{s_1,s_2,0}&=\frac{i^{{s_1}+{s_2}-1} 
\Gamma ({s_1}+{s_2}+1)}{\Gamma (2 {s_1}) \Gamma (2 {s_2}+1)}\ ,
\label{vertex3}
\end{align}
\esubeqs
 for $s_1>s_2$. As we already commented in the Introduction, the vertex above can be used to obtain $0-s_1-s_2$ correlators for $s_1\neq s_2$, but extrapolation to $s_1=s_2$ will give the correct answer too since the correlation function depends smoothly on spins. Nevertheless, the $0-s-s$ correlators should originate from the $CC$-terms in \eqref{UCC}, which previously gave infinite result \cite{Giombi:2009wh,Boulanger:2015kfa}. Therefore, we would like to use the proposal of \cite{Vasiliev:2016xui}, where the new $CC$-terms are
\begin{align}
 \mathcal{U}(h,C,C)&=\frac14\,\int_0^1d\tau\,\int \frac{d\bru d\brv}{(2\pi)^2}\,e^{i\bru_{\gbd}\brv^{\gbd}}\,h_{\ga\gad}y^{\ga}\left[\tau\bru+(1-\tau)\brv\right]^\gad\,C(\tau y,\bry+\bru)C((1-\tau)y,\bry+\brv)\nonumber\\
 &+\text{h.c}\,.
\end{align}
A simple computation along the lines above\footnote{E.Sk is grateful to Massimo Taronna with whom this was obtained last June, see also \cite{Taronna}.} gives\footnote{The issue of $\theta$-dependence is unclear to us since in \cite{Vasiliev:2016xui} it was proposed to take the $\mathcal{N}=2$ supersymmetric HS model and truncate it to the bosonic one using the boundary conditions that contain $\theta$-dependence. Nevertheless, our bosonic truncation which does not rely on imposition of $\theta$-dependent boundary conditions gives the right answer.} 
\begin{align}
(\square -4)\phi &=-4 A K_1K_2 \sum_{s} V_{s,s,0}\ .
\end{align}
Therefore, the $0-s-s$ vertex turns out to have the same form as the naive extrapolation of \eqref{vertexmain} and will give the correct answer without any additional computation needed once \eqref{vertexmain} is shown to be correct.

%%%%%%%%%%%%%%%%%%%%%%%%%%%%%%%%%%%%%%%%%%%%%%%%%%%%%%%%%%%%%%%%%%%%%%%
\section{Computation of the Cubic Amplitude}
\label{sec:amplitude}
%%%%%%%%%%%%%%%%%%%%%%%%%%%%%%%%%%%%%%%%%%%%%%%%%%%%%%%%%%%%%%%%%%%%%%%

From the bulk vertex \eq{vertex2}, the Witten diagram amplitude for $\langle J_{s_1}J_{s_2}J_0\rangle$ 
for $s_1 > s_2$ is obtained as 
\begin{align}
\begin{aligned}
&\langle J_{s_1}(\rmx_1,\eta_1)J_{s_2}(\rmx_2,\eta_2)J_0(\rmx_3)\rangle_{h.s.} 
= (-4A)c_{s_1}c_{s_2} c_0\,  v_{s_1,s_2,0}\times \\
&\qquad\qquad\times\int \frac{d^3\rmx\, dz}{z^4} K_1K_2(K_3)^\Delta(\xi_1 F_2 \bar{\xi}_1)^{s_1-s_2}[(\xi_1\xi_2)^{2s_2}e^{i\theta}+(\bar{\xi}_1\bar{\xi}_2)^{2s_2}e^{-i\theta}]\ ,
\end{aligned}
\label{basic}
\end{align}
where 
\begin{align}
K_i &=K(\rmx-\rmx_i,z)\ , &
F^{\ga\gad}_i &=F^{\ga\gad}(\rmx-\rmx_i,z)\ , &
\xi^{\ga}_i &=\xi^{\ga}(\rmx-\rmx_i,z;\eta_i)\ , &
\bar{\xi}^{\gad}_i &=\bar{\xi}^{\gad}(\rmx-\rmx_i,z;\eta_i) \ .
\end{align}
The factor $(-4A)$ in \eqref{basic} does not have any physical meaning and 
is an arbitrary normalization factor between $\omega$ and $C$ in \eqref{OMSTcocycle}. 
For convenience, we reproduce the factor
\begin{align}
\label{vertexcoef}
v_{s_1,s_2,0}&=\frac{i^{{s_1}+{s_2}-1} 
\Gamma ({s_1}+{s_2}+1)}{\Gamma (2 {s_1}) \Gamma (2 {s_2}+1)}\ .
\end{align}
Also, we introduced normalization factors $c_s$ for each of the three fields. These 
cannot be fixed from the equations of motion and correspond to a freedom on the CFT 
side to normalize at will the two-point functions $\langle j_s j_s\rangle$. Lastly, 
there are two options for boundary conditions on the scalar fields: $\Delta=1$ and $\Delta=2$.

The three-point integrals are doable in principle due to the fact that one can always 
`scalarize' the integrand by representing all $x^{\ga\gad}$-factors as derivatives with 
respect to the boundary points $\rmx_i$. The scalar three-point integral was done long 
ago in \cite{Freedman:1998tz}. The problem is to scalarize in the most efficient way as 
to break as less symmetries as possible. We extend the inversion method of \cite{Freedman:1998tz} 
to our case. Firstly, using the translation invariance we can set $\rmx_1=0$. Then, we apply 
the inversion map both to the boundary and bulk data. As a result the basic structures 
that enter the integrand drastically simplify:
\besubeqs
\begin{align}
\frac{d^3\rmx dz}{z^4}&\rightarrow \frac{d^3\rmx dz}{z^4}\ ,
\\
K_1&\rightarrow z\ ,
\\
K_{2,3}&\rightarrow \rmx_{2,3}^2 K_{2,3}\ ,
\\
(\xi_1 F_2\bar{\xi}_1) &\rightarrow 2 z K_2 [\eta_1 (\rmx-\rmx_2)\eta_1]=-2 z K_2T_{11}\ ,
\\
(\xi_2 F_1\bar{\xi}_2) &\rightarrow 2 K_2^2 [\eta_2 (\rmx-\rmx_2)\eta_2]=-2 K_2^2T_{22}\ ,
\\
(\xi_1\xi_2)+(\bar\xi_1\bar\xi_2)&\rightarrow -2 z K_2 (\eta_1\eta_2)\ ,
\\
(\xi_1\xi_2)-(\bar\xi_1\bar\xi_2)&\rightarrow 2 i K_2 [\eta_1 (\rmx-\rmx_2)\eta_2]=-2 i K_2 T_{12}\ ,
\end{align}
\esubeqs
where we defined
\begin{align}\label{bulktij}
T_{ij}&=-[\eta^i_\ga (\rmx-\rmx_2)^{\ga\gb}\eta^j_\gb]\ ,
\end{align}
and we will use the same notation for the variables after the inversion is applied. 
Our strategy is to rewrite the integrand in terms of simple differential operators 
acting on a scalar integrand 
\be
\int \frac{d^3\rmx dz}{z^4} z^a (K_2)^b (K_3)^\Delta= (\rmx_{23})^{a-b-\Delta} I_{a,b,\Delta}\ ,
\label{mi}
\ee
where
\be
I_{a,b,\Delta}=\frac{\pi ^{3/2} \Gamma \left(\frac{1}{2} (a+b-\Delta )\right) \Gamma 
\left(\frac{1}{2} (a-b+\Delta )\right) \Gamma \left(\frac{1}{2} (-a+b+\Delta )\right) 
\Gamma \left(\frac{1}{2} (a+b+\Delta -3)\right)}{2 \Gamma (a) \Gamma (b) \Gamma (\Delta )}\ .
\ee
There are three operators that can be immediately observed to generate the same 
structures that occur under the integral sign:
\begin{align}
O_{11}&= (\eta_1 \pl_2 \eta_1)\equiv \eta_1^\ga \frac{\pl}{\pl\rmx_2^{\ga\gb}} \eta_1^\gb\ , &
O_{12}&= (\eta_1 \pl_2 \eta_2)\ , &
O_{22}&= (\eta_2 \pl_2 \eta_2) \ .
\end{align}
The operators act on $K_2$ factors only and yield:
\begin{align}
O_{ij} f(K_2)&= \frac{(K_2)^2}{z} \frac{\pl}{\pl K_2} f(K_2) \,T_{ij} \ .   
\end{align}
There is one relation between $T_{ij}$ that is the bulk analog of the 
$S_3^2 +Q_1 Q_2-P_{12}^2\equiv0$ relation:
\begin{align}
(T_{12})^2&= T_{11}T_{22}+\rmx_{23}^2 (\eta_1\eta_2)^2\ .
\end{align}
The integrand can be represented as a function of $O_{ij}$ and $(\eta_1\eta_2)$ 
acting on the scalar integrand, and there is more than one way to do so due to 
the identity above. Then, the integral can be done and one is left with the same 
functional acting on some powers of $|\rmx_{23}|$, which clearly generates some 
function of the conformally invariant structures resulting from setting $\rmx_1=0$ 
followed by the inversion map:
\besubeqs\label{cfti}
\begin{align}
\rmx_{2,3}&\rightarrow \frac{1}{\rmx_{2,3}}\ ,
\\
\rmx_{23}&\rightarrow \frac{\rmx_{23}}{\rmx_2 \rmx_3}\ ,
\\
P_{12}&\rightarrow (\eta_1\eta_2)\ ,
\\
Q_1&\rightarrow -[\eta_1 \rmx_{23}\eta_1]\ ,
\\
Q_2&\rightarrow +\left[\eta_2 {\rmx_{23}}{}\eta_2\right]\frac{1}{\rmx_{23}^2}\ ,
\\
S_3&\rightarrow +\left[\eta_1 {\rmx_{23}}\eta_2\right]\frac{1}{\rmx_{23}}\ .
\end{align}
\esubeqs
It is convenient to use the same notation $T_{ij}$ for the corresponding structures 
on the boundary $T_{ij}=[\eta_i \rmx_{23} \eta_j]$ since they arise under the action 
of $O_{ij}$ on $\rmx_{23}$ resulting from the integral. In other words, if $O_{ij}$ 
applied to the l.h.s. \eqref{mi}, it generates bulk $T_{ij}$ defined in \eqref{bulktij}, 
and if $O_{ij}$ acts on the r.h.s. of \eqref{mi}, it generates the boundary $T_{ij}$.

%%%%%%%%%%%%%%%%%%%%%%%%%%%%%%%%%%%%%%%%%%%%%%%%%%%%%%%%%%%%%%%%%%%%%%%
\subsection{Leading Coefficients}
%%%%%%%%%%%%%%%%%%%%%%%%%%%%%%%%%%%%%%%%%%%%%%%%%%%%%%%%%%%%%%%%%%%%%%%

Our goal is to reproduce the full structure of the three-point functions. However, 
it is useful to perform a few simple checks of the duality that do not require establishing 
a full dictionary between bulk and boundary. It is clear that $O_{ij}$ operators when 
applied one after another produce 
\begin{align}
\label{leadcoef}
O_{11}^{n_1}O_{12}^{n_2}O_{22}^{n_3} (K_2)^a&= \frac{ \Gamma[a+n]}{z^n\Gamma[a]}(K_2)^{a+n}
T_{11}^{n_1}T_{12}^{n_2}T_{22}^{n_3} + {\cal O}(\eta_1\eta_2)\ ,
\end{align}
where $n=n_1+n_2+n_3$. Therefore, up to $P_{12}$-terms, which are represented by 
${\cal O}(\eta_1\eta_2)$ after the inversion, the bulk computation amounts to pulling out 
the $T_{ij}$ structures as powers of $O_{ij}$, computing the scalar integral and then 
pushing the $O_{ij}$ factors in. In the last step operators $O_{ij}$ act on $\rmx_{23}$ 
resulting from the bulk integral and generate $Q_{1,2}$ modulo $Q_1Q_2 \sim -S_3^2$. 

\paragraph{Type-A, $\boldsymbol{\Delta=1}$.} In the case of $\Delta=1$ and $\theta=0$ 
a simple computation gives:
\begin{align}
\langle J_{s_1}(\rmx_1,\eta_1)J_{s_2}(\rmx_2,\eta_2)J_0(\rmx_3)\rangle_{h.s.} &= 
(-4A)c_{s_1}c_{s_2} c_0\,  v_{s_1,s_2,0} I^{\Delta=1}_{s_1,s_2,0}\times 
\left[ (Q_1)^{s_1} (Q_2)^{s_2}+ {\cal O}(P_{12})\right] \ ,
\end{align}
where $I_{s_1,s_2,0}$ is the factor that comes from the bulk 
\begin{align}
I^{\Delta=1}_{s_1,s_2,0}=\frac{\pi ^3 \left(-\frac{1}{2}\right)^{s_1+s_2} 
(-)^{s_1}\Gamma \left(s_1+\frac{1}{2}\right)}{s_1 \Gamma 
\left(\frac{1}{2}-s_2\right) \Gamma (s_1+s_2+1)}\ .
\end{align}
This should be compared with the generating function \eqref{bosonjjo} in free boson theory:\footnote{What we compute would correspond to the dual theory with $U(M)$ global symmetry \eqref{bosonjjoU}. The same comment applies to  all correlators below.}
\begin{align}
\langle J_{s_1}(\rmx_1,\eta_1)J_{s_2}(\rmx_2,\eta_2)J_0(\rmx_3)\rangle_{f.b.} &=\frac{2}{\rmx_{12}\rmx_{23}\rmx_{13}}
\left(\frac{i}{2}\right)^{s_1+s_2}\frac{1}{s_1! s_2!}\left[(Q_1)^{s_1} (Q_2)^{s_2}
+ {\cal O}(P_{12})\right]\ .
\end{align}
Since the normalization of the boundary-to-bulk propagators is not yet fixed, we should 
compare two function of $s_1,s_2$, the bulk result and the CFT result, up to a product of 
functions of $s_1$ and $s_2$ separately. In doing so we can find that the normalization factor is
\begin{align}
c_s&=-\frac{4^s}{\pi }\ .
\end{align}
With this normalization we have a perfect match up to the terms of order $(\eta_1\eta_2)=P_{12}$, namely
\begin{align}
\langle J_{s_1}(\rmx_1,\eta_1)J_{s_2}(\rmx_2,\eta_2)J_0(\rmx_3)\rangle_{h.s.}=\langle J_{s_1}(\rmx_1,\eta_1)J_{s_2}(\rmx_2,\eta_2)J_0(\rmx_3)\rangle_{f.b.}
+{\cal O}(P_{12})\ .
\end{align}

\paragraph{Type-B, $\boldsymbol{\Delta=2}$.} The same computation but for $\Delta=2$ 
and $\theta=\pi/2$ leads to
\begin{align}
\langle J_{s_1}(\rmx_1,\eta_1)J_{s_2}(\rmx_2,\eta_2)J_0(\rmx_3)\rangle_{h.s.} 
&= (-4A)c_{s_1}c_{s_2} \tilde{c}_0\,  v_{s_1,s_2,0} I^{\Delta=2}_{s_1,s_2,0}\times 
\left[ (Q_1)^{s_1-1} (Q_2)^{s_2-1}P_{12}S_3+ {\cal O}(P_{12}^2)\right] \ .
\end{align}
Here we assumed that the normalization of the scalar field propagator $\tilde{c}_0$ can 
be different from $c_0$ due to the change in boundary conditions. The bulk integral and 
other factors combine into $I^{\Delta=2}_{s_1,s_2,0}$ as follows
\begin{align}
I^{\Delta=2}_{s_1,s_2,0}=-\frac{\pi ^3 s_2 \left(-\frac{1}{2}\right)^{s_1+s_2-1} (-)^{s_1} 
\Gamma \left(s_1+\frac{1}{2}\right)}{\Gamma \left(\frac{1}{2}-s_2\right) \Gamma (s_1+s_2+1)}\ .
\end{align}
We note that one factor of $P_{12}$ jumps out of the integral and the subleading terms are 
of order $P_{12}^2$. The result should be compared with the generating function 
\eqref{fermionjjo} in free fermion theory:
\begin{align}
\langle J_{s_1}(\rmx_1,\eta_1)J_{s_2}(\rmx_2,\eta_2)J_0(\rmx_3)\rangle_{f.f.} &=
\frac{i^{s_1+s_2-2}}{\rmx_{23}^2\rmx_{13}^22^{s_1+s_2-2}}\frac{2}{(s_1-1)! (s_2-1)!}
\left[(Q_1)^{s_1-1} (Q_2)^{s_2-1}S_3P_{12}+ {\cal O}(P_{12}^2)\right]\ .
\end{align}
We find the two results to agree, namely
\begin{align}
\langle J_{s_1}(\rmx_1,\eta_1)J_{s_2}(\rmx_2,\eta_2)J_0(\rmx_3)\rangle_{h.s.}
=-\frac{\tilde{c}_0}{c_0}\langle J_{s_1}(\rmx_1,\eta_1)J_{s_2}(\rmx_2,\eta_2)J_0
(\rmx_3)\rangle_{f.f.}+ {\cal O}(P_{12})\ .
\end{align}
Once the normalization factors $c_s$ are fixed by the Type-A duality the match just 
observed is even more nontrivial because the only freedom that we have is an overall 
spin-independent factor.\footnote{The generating functions of the HS currents built out 
of free bosons and fermions, which were introduced in Section \ref{sec:CFT}, are components 
of the supermultiplet of HS currents. Therefore, the normalization of the HS currents in free 
boson and free fermion are naturally related to each other. } The canonical normalization of the scalar 
field propagators \cite{Freedman:1998tz} is such that $\tilde{c}_0/c_0=-2$. 

%%%%%%%%%%%%%%%%%%%%%%%%%%%%%%%%%%%%%%%%%%%%%%%%%%%%%%%%%%%%%%%%%%%%%%%
\subsection{Complete Dictionary}
%%%%%%%%%%%%%%%%%%%%%%%%%%%%%%%%%%%%%%%%%%%%%%%%%%%%%%%%%%%%%%%%%%%%%%%

We would like to reproduce the full structure of CFT correlators. The idea is to take the 
subleading terms into account and express the bulk integrand as the action of a differential 
operator in $O_{ij}$ on the scalar integrand. The operators $O_{ij}$ produce $T_{ij}$ and 
the subleading terms are obtained due to
\begin{align}
O_{11}T_{22}&=(\eta_1\eta_2)^2\ , &
O_{22}T_{11}&=(\eta_1\eta_2)^2\ , &
O_{12}T_{12}&=-\frac12(\eta_1\eta_2)^2 \ .
\end{align}
Operators $O_{ij}$ commute with each other. Moreover, $O_{11}$ does not produce any 
subleading terms at all. Then the action of any power of, say $O_{12}$, can be evaluated 
starting from ($\rmx\equiv\rmx_{23}$)
\begin{align}
O_{12}f(\rmx^2, T_{12}) &= \left(- T_{12}\frac{\pl}{\pl \rmx^2} 
-\frac12 (\eta_1\eta_2)^2\frac{\pl}{\pl T_{12}}\right) f(\rmx^2,T_{12})\ ,
\end{align}
and exponentiating it as
\begin{align}
\exp[tO_{12}]\, (\rmx^2)^{-a}&= \left(\rmx^2 -t T_{12} +\frac{t^2}{4} (\eta_1\eta_2)^2\right)^{-a}\ ,
\end{align}
which leads to Gegenbauer polynomials: 
\begin{align}
\label{directform}
(O_{12})^n (\rmx^2)^{-a}&=\sum_k A^{a,n}_k(\eta_1\eta_2)^{2k} 
(T_{12})^{n-2k} (\rmx^2)^{-(a+n-k)}\,,&& A^{a,n}_k=\frac{\left(-\right)^k n! 
\Gamma (a-k+n)}{4^kk! \Gamma (a) (n-2 k)!}\ .
\end{align}
In particular, we find \eqref{leadcoef}
\begin{align}
(O_{12})^n (\rmx^2)^{-a}&= \frac{\Gamma[a+n]}{\Gamma[a]} 
(\rmx^2)^{-(a+n)}(T_{12})^n+ {\cal O}(\eta_1\eta_2)\ .
\end{align}
In fact, we will use the inversion formula:
\begin{align}\label{invform}
\sum_{k=0}B^{a,n}_k(\eta_1\eta_2)^{2k}(O_{12})^{n-2k} (\rmx^2)^{-(a+k-n)}= 
(\rmx^2)^{-a}(T_{12})^n\ , && B^{a,n}_k= 
\frac{\Gamma[a+k-n]n!}{\Gamma[a]4^k k!(n-2k)!}(\eta_1\eta_2)^{2k}\ .
\end{align}
The same formula works in the bulk if we replace $|\rmx^2|^{-1}$ by $K_2/z$ and $T_{ij}$ 
by the bulk $T_{ij}$ \eqref{bulktij} which we happen to denote  by the same symbol, as explained below
\eqref{cfti}. Therefore, any  $T_{12}$ structure can also be factored out.

%%%%%%%%%%%%%%%%%%%%%%%%%%%%%%%%%%%%%%%%%%%%%%%%%%%%%%%%%%%%%%%%%%%%%%%
\subsection{Complete Three-Point Functions}
\label{subsec:complete}
%%%%%%%%%%%%%%%%%%%%%%%%%%%%%%%%%%%%%%%%%%%%%%%%%%%%%%%%%%%%%%%%%%%%%%%

We can now compute \eqref{basic} for general $\theta$. The choice $\Delta=1$ 
should be compared with CS-boson and $\Delta=2$ with CS-fermion.

The computation is now reduced to the following simple steps: (i) express the integrand 
in terms of $(\xi_1\xi_2)\pm (\bar\xi_1\bar\xi_2)$; (ii) apply the inversion map and 
express the integrand in terms of $T_{ij}$; (iii) $T_{11}$ is factored out immediately, 
while $T_{12}$ is pull out with the help of \eqref{invform} as certain polynomial in 
$O_{12}$; (iv) the integral can be done; (v) the operators $O_{ij}$ should be evaluated 
in terms of $T_{ij}$, which is easy for $O_{11}$ and we use \eqref{directform} for $O_{12}$; 
(vi) powers of $T_{ij}$ should be replaced by $Q_1$, $Q_2$, $P_{12}$, $S_3$.

As the first step we need the following simple identity, which reveals the dependence 
on $\theta$ of the final result as well as how it splits into parity-even and 
parity-odd structures:
\begin{align}
\label{thetabulk}
\begin{aligned}
[(\xi_1\xi_2)^{2s_2}e^{i\theta}+(\bar{\xi}_1\bar{\xi}_2)^{2s_2}e^{-i\theta}]&=
\sum_k \frac{2\cos \theta}{4^{s_2}}C^{2s_2}_{2k}[(\xi_1\xi_2)+ 
(\bar\xi_1\bar\xi_2)]^{2s_2-2k}[(\xi_1\xi_2)- (\bar\xi_1\bar\xi_2)]^{2k}+
\\
&+\sum_k \frac{2i\sin \theta}{4^{s_2}}C^{2s_2}_{2k+1}[(\xi_1\xi_2)+ 
(\bar\xi_1\bar\xi_2)]^{2s_2-2k-1}[(\xi_1\xi_2)- (\bar\xi_1\bar\xi_2)]^{2k+1}\ .
\end{aligned}
\end{align}
Since the $\theta$-dependence is fixed, in computing the correlation functions it is sufficient to choose either $\theta=0$ or $\theta=\pi/2$ and for each of these cases to consider the $\Delta=1$ and $\Delta=2$ boundary conditions for the scalar field.  Therefore,  we need to compute four bulk integrals. In fact, these terms 
can be computed in the parity even Type-A,B theories but with different choice of boundary 
conditions for the scalar field, see also the end of Section \ref{subsec:csmatter}.

The comment that applies to all the cases considered below is that the generating 
functions of correlators depend smoothly on spins and therefore it should be possible 
to extrapolate the result in the case of $s_1>s_2$ to the case of $s_1=s_2$, which is 
an argument used in \cite{Giombi:2009wh}. The assumption $s_1>s_2$ was used to pick only 
$\omega_{s_1}C_{s_2}$ terms and for the opposite situation we find contribution from 
$\omega_{s_2}C_{s_1}$, which is the same. Therefore, the computation below covers all 
possible $s_1$ and $s_2$.

Another comment is that the results we obtain in the bulk are valid for the case of $U(N)$ CFT's that have HS currents with all integer spins (there are only HS currents with even spin in the $O(N)$ case). Also, the bulk results are valid for the case of the leftover global symmetry on the CFT side. The projection onto the singlet sector is trivial and can be obtained by 
taking the bose symmetric part of the correlator, as discussed in Section \ref{subsec:vertices}. It should be noted that the result of \cite{Giombi:2016zwa}, which we will compare the AdS/CFT correlators with, were obtained assuming that the prediction of the slightly broken HS symmetry \cite{Maldacena:2012sf} extends to all integers spins and possibly to the case of leftover global symmetries. 

%%%%%%%%%%%%%%%%%%%%%%%%%%%%%%%%%%%%%%%%%%%%%%%%%%%%%%%%%%%%%%%%%%%%%%%
\subsubsection{Type-A, Free Boson}
%%%%%%%%%%%%%%%%%%%%%%%%%%%%%%%%%%%%%%%%%%%%%%%%%%%%%%%%%%%%%%%%%%%%%%%

The integral corresponding to $\Delta=1$ and $\theta=0$ is given by
\begin{align}
\label{deltaoneeven}
\begin{aligned}
&\langle J_{s_1}(\rmx_1,\eta_1)J_{s_2}(\rmx_2,\eta_2)J_0 
(\rmx_3)\rangle_{h.s.}^{\Delta=1,\theta=0} = (-4A)c_{s_1}c_{s_2} c_0\,  v_{s_1,s_2,0}\times 
\\
&\quad\times\int \frac{d^3\rmx dz}{z^4} K_1K_2K_3(\xi_1 F_2 \bar{\xi}_1)^{s_1-s_2}
\sum_k \frac{2}{4^{s_2}}C^{2s_2}_{2k}[(\xi_1\xi_2)+ (\bar\xi_1\bar\xi_2)]^{2s_2-2k}
[(\xi_1\xi_2)- (\bar\xi_1\bar\xi_2)]^{2k}\ .
\end{aligned}
\end{align}
The final result for the right hand side of this equation is the following triple sum
\begin{align}
\sum_{k=0}^{s_2}\sum_{i=0}^{k}\sum_{j=0}^{k-i}&\frac{(-1)^{k-j}  i^{s_1+s_2} 2^{s_1-s_2+1}   
\Gamma (2 (-k+s_1+s_2))  \Gamma \left(-i+k+s_1-s_2+\frac{1}{2}\right) }{\sqrt{\pi }  
i!j!  (2 s_1-1)! (2 s_2-2k)!  (-i+2 s_1)!  (-i-j+k)!}
\nonumber\\ 
&\times \frac{Q_1^{s_1-s_2}P_{12}^{2 (i+j-k+s_2)}
\left(P_{12}^2-Q_1 Q_2\right)^{-i-j+k}}{\rmx_{12}\rmx_{13}\rmx_{23}}\ ,
\label{ts}
\end{align}
where we undid the inversion map. This expression can be supplemented with 
the $s_2>s_1$ contribution, extended to diagonal $s_1=s_2$, as we explained above, 
then summed over spins as to build a generating function, see Appendix \ref{app:freeboson}, 
giving the result:
\be
\frac{2}{\rmx_{12}\rmx_{13}\rmx_{23}}\exp{\left(\tfrac{i}2{Q_1}+\tfrac{i}2 Q_2\right)}
\cos P_{12}\ .
\ee
This matches exactly the CFT three-point function:
\begin{align}
\langle J_{s_1}(\rmx_1,\eta_1)J_{s_2}(\rmx_2,\eta_2)J_0
(\rmx_3)\rangle_{h.s.}^{\Delta=1,\theta=0}=\langle J_{s_1}(\rmx_1,\eta_1)
J_{s_2}(\rmx_2,\eta_2)J_0(\rmx_3)\rangle_{f.b.}\ .
\end{align}
This result is slightly new as we managed to reproduce the full structure of the CFT 
correlators from the local HS equations where the field theory methods can 
be applied, and we note that no regularization of infinities is required. The leading coefficients 
were found in \cite{Giombi:2009wh} and the same generating function resulted from the 
non-local $CC$-terms after some regularization in \cite{Giombi:2010vg}.

%%%%%%%%%%%%%%%%%%%%%%%%%%%%%%%%%%%%%%%%%%%%%%%%%%%%%%%%%%%%%%%%%%%%%%%
\subsubsection{Type-B, Free Fermion}
%%%%%%%%%%%%%%%%%%%%%%%%%%%%%%%%%%%%%%%%%%%%%%%%%%%%%%%%%%%%%%%%%%%%%%%

The integral corresponding to $\Delta=2$ and $\theta=\pi/2$ reads
\begin{align}
\begin{aligned}
&\langle J_{s_1}(\rmx_1,\eta_1)J_{s_2}(\rmx_2,\eta_2)J_0(\rmx_3)\rangle_{h.s.}^{\Delta=2,\theta=\tfrac{\pi}{2}
} = (-4A)c_{s_1}c_{s_2} \tilde c_0\,  v_{s_1,s_2,0}\times 
\\
&\quad\times\int \frac{d^3\rmx dz}{z^4} K_1K_2K_3^2(\xi_1 
F_2 \bar{\xi}_1)^{s_1-s_2}\sum_k \frac{2i}{4^{s_2}}C^{2s_2}_{2k+1}[(\xi_1\xi_
2)+ (\bar\xi_1\bar\xi_2)]^{2s_2-2k-1}[(\xi_1\xi_2)- (\bar\xi_1\bar\xi_2)]^{2k+1}\ .
\end{aligned}
\end{align}
The right hand side of this equation can be brought to the form 
\begin{align}
\frac{\tilde{c}_0}{c_0}\sum_{k=0}^{s_2-1}\sum_{i=0}^{k}\sum_{j=0}^{k
-i}&\frac{(-1)^{j+k} 2^{s_1-s_2+1} i^{s_1+s_2} \Gamma (-2 k+2 s_1+2 s_2
-1) \Gamma \left(-i+k+s_1-s_2+\frac{3}{2}\right)}{\sqrt{\pi } i! j! \Gamma (2 s
_1) \Gamma (-i+2 s_1+1) \Gamma (2 s_2-2 k) \Gamma (-i-j+k+1)} 
\non\\ 
&\times \frac{S_3 Q_1^{s_1-s_2} P_{12}^{2 i+2 j-2 k+2 s_2
-1} \left(P_{12}^2-Q_1Q_2\right)^{-i-j+k}}{\rmx^2_{23}\rmx_{13}^2}\ .
\label{ts2}
\end{align}
As before, this expression can be supplemented with 
the $s_2>s_1$ contribution, extended to diagonal $s_1=s_2$,  
then summed over spins as to build a generating function, see Appendix \ref{app:freefermion}, 
giving the result:
\begin{align}
-\frac{\tilde{c}_0}{c_0}\frac{\exp(\tfrac{i}2 Q_1+\tfrac{i}2 Q
_2)}{\rmx^2_{23}\rmx_{13}^2}S_3 \sin P_{12}\ ,
\end{align}
and matches exactly the CFT three-point function \eqref{freefermjooU}:\footnote{We again note that the answer is formally for the free fermion with leftover $U(M)$ global symmetry.}
\begin{align}
\langle J_{s_1}(\rmx_1,\eta_1)J_{s_2}(\rmx_2,\eta_2)J_0(\rmx_3)\rangle_{h.s.}^{\Delta=2,\theta=\tfrac{\pi}{2}}=\langle J_{s_1}(\rmx_1,\eta_1)J_{s_2}(\rmx_2,\eta_2)J_0(\rm
x_3)\rangle_{f. f.}\ .
\end{align}
where we recall that the prefactor accounts for the difference in the 
boundary conditions for the scalar field and is equal to $-2$.

%%%%%%%%%%%%%%%%%%%%%%%%%%%%%%%%%%%%%%%%%%%%%%%%%%%%%%%%%%%%%%%%%%%%%%%
\subsubsection{Type-A, Critical Boson}
%%%%%%%%%%%%%%%%%%%%%%%%%%%%%%%%%%%%%%%%%%%%%%%%%%%%%%%%%%%%%%%%%%%%%%%

We take the $\Delta=1$ Type-A expression \eqref{deltaoneeven} and simply set $\Delta=2$. This gives
\begin{align}
\begin{aligned}
&\langle J_{s_1}(\rmx_1,\eta_1)J_{s_2}(\rmx_2,\eta_2)J_0(\rmx_3)\rangle_{h.s.}^{\Delta=2,\theta=0} = (-4A)c_{s_1}c_{s_2} \tilde c_0\,  
v_{s_1,s_2,0}\times 
\\
&\quad\times\int \frac{d^3\rmx dz}{z^4} K_1K_2K_3^2(\xi_1
 F_2 \bar{\xi}_1)^{s_1-s_2}\sum_k \frac{2}{4^{s_2}}C^{2s_2}_{2k}[(\xi_1\xi_2)+ (\bar\xi_1\bar\xi_2)]^{2s_2-2k}[(\xi_1\xi_2)- (\bar\xi_1\bar\xi_2)]^{2k}\ .
\end{aligned}
\end{align}
The right hand side of this equation is evaluated to yield
\begin{align}
\label{tAd2sum}
\frac{\tilde{c}_0}{c_0}\sum_{k=0}^{s_2}\sum_{i=0}^{k}
\sum_{j=0}^{k-i}&\frac{(-1)^{k-j} 2^{s_1-s_2+1} i^{s_1+s_2} 
\Gamma (-2 k+2 s_1+2 s_2) \Gamma (-i+k+s_1-s_2+1)}{\sqrt{\pi } i! j! 
\Gamma (2 s_1) \Gamma (-i+2 s_1+1) \Gamma (-2 k+2 s_2+1) \Gamma 
\left(-i-j+k+\frac{1}{2}\right)}
\non
\\ 
&\times \frac{Q_1^{s_1-s_2} 
P_{12}^{2 (i+j-k+s_2)} \left(P_{12}^2-Q_1Q_2\right)^{-i-j+k}}{\rmx^2_{23}\rmx_{13}^2}\ .
\end{align}
The three-point function in the CS-fermion theory was found in a form of recurrence relations 
in \cite{Giombi:2016zwa} (see also Section \ref{subsec:criticalboson}). We find that the triple sum above 
is an explicit solution to those recursion relations. The leading coefficient
is
\begin{align}
\langle J_{s_1}(\rmx_1,\eta_1)J_{s_2}(\rmx_2,\eta_2)J_0(\rmx_3)\rangle_{h.s.}^{\Delta=2,\theta=0}&=-\frac{1}{2}\frac{\tilde{c}_0}
{c_0}\frac{i^{s_1+s_2} 2^{-s_1-s_2+2}}{\Gamma \left(s_1+\frac{1}{2}\right) \Gamma \left(s_2+\frac{1}{2}\right)}\frac{Q_1^{s_1}Q_2^{s_2}}{\rmx^2_{23}\rmx_{13}^2}+ {\cal O}(P_{12})\ .
\end{align}
The sum \eqref{tAd2sum} can be evaluated explicitly, see Appendix \ref{app:criticalboson}. The computation shows that it is more convenient to use the $S_3$ variable. Indeed, the bulk integral gives $S_3$ after $O_{12}$ is applied. We expect this to be the case for the most general case of three non-zero spins too: it may be more convenient to express even structures in terms of even functions of the odd $S$ structures.

%%%%%%%%%%%%%%%%%%%%%%%%%%%%%%%%%%%%%%%%%%%%%%%%%%%%%%%%%%%%%%%%%%%%%%%
\subsubsection{Type-B, Critical Fermion}
%%%%%%%%%%%%%%%%%%%%%%%%%%%%%%%%%%%%%%%%%%%%%%%%%%%%%%%%%%%%%%%%%%%%%%%

Using the same technique as before we can compute the parity odd integral
 for $\Delta=1$, which for $\Delta=2$ reproduced the free fermion result, the 
only modification required being $K_3^2\rightarrow K_3$:
\begin{align}
\begin{aligned}
&\langle J_{s_1}(\rmx_1,\eta_1)J_{s_2}(\rmx_2,\eta_2)J_0(\rmx_3)
\rangle_{h.s.}^{\Delta=1,\theta=\tfrac{\pi}{2
}} = (-4A)c_{s_1}c_{s_2} c_0\,  v_{s_1,s_2,0}\times 
\\
&\quad\times\int \frac{d^3\rmx dz}{z^4} K_1K
_2K_3(\xi_1 F_2 \bar{\xi}_1)^{s_1-s_2}\sum_k 
\frac{2i}{4^{s_2}}C^{2s_2}_{2k+1}[(\xi_1\xi_2)+ 
(\bar\xi_1\bar\xi_2)]^{2s_2-2k-1}[(\xi_1\xi_2)- (\bar\xi_1\bar\xi_2)]^{2k+1}\ .
\end{aligned}
\end{align}
As a result, the right hand side of this equation yields the following  triple sum representation 
for the critical fermion:
\begin{align}
\label{tbd1sum}
\sum_{k=0}^{s_2-1}\sum_{i=0}^{k}
\sum_{j=0}^{k-i}&\frac{(-1)^{j+k} i^{s_1+s_2} 2^{s_1-s_2+1} 
\Gamma (-2 k+2 s_1+2 s_2-1) \Gamma (-i+k+s_1-s_2+1)}{\sqrt{\pi } i! j! 
\Gamma (2 s_1) (-2 k+2 s_2-1)! \Gamma (-i+2 s_1+1) \Gamma \left(-i-j+k+\frac{3}{2}
\right) }\non
\\ 
&\times \frac{S_3 Q_1^{s_1-s_2} P_{12}^{2 i+2 j-2 k+2 s_
2-1} \left(P_{12}^2-Q_1 Q_2\right)^{-i-j+k}}{\rmx_{12}\rmx_{13}\rmx_{23}}\ .
\end{align}
In \cite{Giombi:2016zwa}, see also Section \ref{subsec:criticalfermion}, 
the generating function of the correlators was found in CS-boson theory in an implicit form 
of a recurrence relation. The triple sum above provides an explicit solution to this 
system. The leading coefficient is easy to find, and gives
\begin{align}
\langle J_{s_1}(\rmx_1,\eta_1)J_{s_2}(\rmx_2,\eta_2)J_0 
(\rmx_3)\rangle_{h.s.}^{\Delta=1,\theta=\tfrac{\pi}{2}} &= 
-\frac{i^{s_1+s_2} 2^{-s_1-s_2+2}}{\Gamma \left(s_1+\frac{1}{2}\right) 
\Gamma \left(s_2+\frac{1}{2}\right)}\frac{Q_1^{s_1-1}Q_2^{s_2-1}S_3 P_
{12}}{\rmx_{12}\rmx_{23}\rmx_{13}}+ {\cal O}(P_{12}^2)\ .
\end{align}
The sum \eqref{tbd1sum} can be evaluated explicitly, see Appendix \ref{app:criticalfermion}.

%%%%%%%%%%%%%%%%%%%%%%%%%%%%%%%%%%%%%%%%%%%%%%%%%%%%%%%%%%%%%%%%%%%%%%%
\subsubsection{Summary}
%%%%%%%%%%%%%%%%%%%%%%%%%%%%%%%%%%%%%%%%%%%%%%%%%%%%%%%%%%%%%%%%%%%%%%%

Combining all the four cases together with the $\theta$ dependence \eqref{thetabulk} we have confirmed that the structure of $\langle j_{s_1} j_{s_2} j_0\rangle$ 
is in accordance with the CFT result \eqref{jjo}:
\besubeqs
\label{jjoB}
\begin{align}
\langle j_{s_1} j_{s_2} j_0\rangle_{B.}&= {\tilde{N}} \left[
\cos\theta \langle j_{s_1} j_{s_2} {j}_0\rangle_{f.b.}+\sin \theta\langle j_{s_1} j_{s_2} j_0
\rangle_{odd}\right]\ ,
\\
\langle j_{s_1} j_{s_2} \tilde{j}_0\rangle_{F.}& = 
{ \tilde{N}} \left[\cos\theta \langle j_{s_1} j_{s_2} \tilde{j}_0
\rangle_{f.f.}+\sin \theta\langle j_{s_1} j_{s_2} \tilde{j}_0\rangle_{odd}\right]\ .
\end{align}
\esubeqs
Here we restored the bulk coupling constant $G=1/\tilde{N}$. The $\theta$-dependence 
results from the fact that the (anti)-selfdual HS Weyl tensors are 
identified with the order $s$-curls of the Fronsdal field with a phase shift 
\eqref{OMSTcocycle}
\besubeqs
\begin{align} 
\label{weyldefinition}
C_{\ga_1...\ga_{2s}} &=e^{+i\theta}\,\nabla\fdu{\ga_1}{\gad_1}...
\nabla\fdu{\ga_s}{\gad_s} \phi_{\ga_{s+1}...\ga_{2s},\gad_1...\gad_s}\ , \\
C_{\gad_1...\gad_{2s}} &=e^{-i\theta}\,\nabla\fud{\ga_1}{\gad
_1}...\nabla\fud{\ga_s}{\gad_s} \phi_{\ga_1...\ga_s,\gad_{s+1}...\gad_ {2s}}\ .
\end{align}
\esubeqs
Given the $\theta$-dependence, the four structures in \eqref{jjoB} can be
 found by evaluating the bulk integral for $\Delta=1,2$ and $\theta=0,\pi/2$, as we did.

%%%%%%%%%%%%%%%%%%%%%%%%%%%%%%%%%%%%%%%%%%%%%%%%%%%%%%%%%%%%%%%%%%%%%%%
\section{Discussion}
\label{sec:discussion}
%%%%%%%%%%%%%%%%%%%%%%%%%%%%%%%%%%%%%%%%%%%%%%%%%%%%%%%%%%%%%%%%%%%%%%%

We have extracted the full structure of the correlators that follow from the 
$\omega C$-correction to the scalar field equation: backreaction of $s_
1$ and $s_2$ onto the $s=0$ field. The perfect agreement with the CFT results is 
found, including the $\theta$-dependence and the parity-violating structures, which 
were recently derived in \cite{Giombi:2016zwa} on the CFT side.

The most general bulk result includes the case of the gauged $U(N)$ symmetry with the 
leftover $U(M)$ global symmetry on the CFT side and should correspond to an extension of  
\cite{Maldacena:2012sf} to these cases. The results allow us to fix a part of the cubic action $s_1-
s_2-0$ that is $\theta$-dependent. Indeed, upon treating various components of 
$\omega$ and $C$ as a bookkeeping device for the derivatives of the Fronsdal 
fields, the vertex has a rather simple form (schematically):
\begin{align}
V=v'_{s_1,s_2,0}\int \phi\, \omega^{\ga(s_1+s_2-1),\gad(s_1-s
_2-1)} C_{\ga(2s_2+s_1-s_2),\gad(s_1-s_2)}\hat{h}^{\ga\gad} +h.c.\,,
\end{align}
where $\hat{h}^{\ga\gad}$ is the basis three-form. This expression, after plugging 
the propagators in place of $\omega$ and $C$ results in \eqref{basic}, i.e. the integral 
we have computed. The spin-dependent part of the coupling $v'_{s_1,s_2,0}$ is fixed by 
the HS algebra structure constants via $\omega\star C-C\star \tilde{\omega}$. 
Therefore, this coupling does not require anything beyond the HS algebra, 
in particular it does not rely on the Vasiliev equations, though they can, 
of course, be derived from them as well. The coupling is local and 
for that reason is easy to deal with.

The gauge invariance of the action or equations of motion implies that the correlation 
functions are those of conserved currents, see e.g. \cite{Mikhailov:1984cp, Sleight:2016xqq}. This is the same statement as the invariance 
of the $S$-matrix of massless fields under the free gauge transformations. The bulk proof 
of this result heavily relies upon neglecting possible boundary terms. However, the CFT 
correlators are made of several structures: those given by free theories do correspond 
to conserved currents, while the parity-odd structures are not conserved when the triangle 
inequality is violated, e.g. $s_1+s_2<s_3$. The non-conservation of a current can be accounted 
for only by non-vanishing boundary terms \cite{Giombi:2011rz} in the approach based on action 
principle. Therefore, the agreement for the parity-violating correlators cannot be explained by 
simple gauge invariance arguments and should be regarded as a test of the duality. Another result \cite{Giombi:2011ya,Bekaert:2012ux} is that the duality with the critical boson/fermion\footnote{In \cite{Giombi:2011ya} the duality between Type-A HS theory and free/critical boson was studied, the extension to Type-B vs. free/critical fermion should be straightforward.} should follow order by order in $1/N$ expansion from the duality with the free boson/fermion. While for $\theta=0,\pi/2$ this implies that the bulk results for critical models are not independent from those for the free models, it is unclear whether this argument can be extended to the full range of $\theta$ as on the CFT side the dependence on $\theta$ results from a highly nontrivial computation \cite{Maldacena:2012sf}.

In this paper we could not extract the correlators for general three spins due to non-localities 
present in the additional $CC$-terms \cite{Giombi:2009wh,Giombi:2010vg,Boulanger:2015ova} that 
need to be taken into account. These non-localities make the coefficient of the bulk vertex 
infinite, resulting in infinite correlators. The infinity is due to the presence of infinitely 
many of higher derivative copies of the same vertex that are stacked in the $CC$-terms. 
The divergences at tree level must not arise in any field theory, 
including HS theories.\footnote{There exist extremal correlators and the divergences observed in 
\cite{Giombi:2009wh,Giombi:2010vg,Boulanger:2015ova} have nothing to do with those. }
Indeed, the correlation functions on the CFT side are finite. Moreover, there is a one-to-one correspondence 
between all possible three-point correlators and cubic vertices in $AdS$, which allows one 
to manufacture an action up to the cubic terms that will yield any given three-point correlation 
functions and no infinities can arise. This was explicitly done for some of the cubic vertices in \cite{Bekaert:2014cea,Skvortsov:2015pea,Sleight:2016dba} (full cubic action of Type-A in any $d$ was obtained in \cite{Sleight:2016dba}), see also \cite{Bekaert:2015tva,Sleight:2017pcz,Sleight:2017fpc} for 
the quartic results. In the recent paper \cite{Vasiliev:2016xui} it was conjectured how to 
modify the $CC$-terms as to make them local. The field redefinition that does the job is non-local 
and changes the coefficient of the correlator from an infinite number to a finite one. Moreover, the 
$CC$-terms correspond to the abelian vertices at this order (in the sense that are not fixed by the HS symmetry)  
and can in principle be arbitrary. Also, non-local redefinitions can result in any given coefficient \cite{Barnich:1993vg,Prokushkin:1999xq,Kessel:2015kna,Boulanger:2015ova,Taronna:2016xrm}. 
Nonetheless, the proposal of \cite{Vasiliev:2016xui} involves a specific type of field redefinition, 
and as such it can be treated as a conjecture. Taking this point of view, we have tested this redefinition, 
which involves the $CC$ terms, and we have found that it does produce an answer for the spin $0-s-s$ amplitude 
that agrees with the CFT result. Further tests of this conjecture will require the study of more three-point functions,
namely those involving three arbitrary spins, and then of higher order amplitudes,
which may possibly require higher order extension of the proposed field redefinition as well.

It is interesting that the $\theta$ dependence enters the Vasiliev equations in a very simple 
way --- $e^{\pm i\theta}$ factors in two terms in the master equations. It was proved 
in \cite{Sharapov:2017yde} that the $\theta$-deformation is indeed unique in HS gravity. In the perturbation theory 
this leads to some proliferation of $\theta$-dependent terms, so that in the $\omega$-equation of 
motion one generally finds terms from $e^{-in\theta }$ to $e^{+in\theta }$ at the $n$-th order. 
Some of them are too non-local to be treated by field theory methods. Hopefully, there are simple 
enough non-local redefinitions that render the couplings finite and do not introduce any spin-dependent 
$\theta$-coefficients. This would imply that the planar $n$-point correlation 
functions of HS currents in Chern-Simons matter theories have a form similar to those of
the bulk three-point amplitudes even for $n>3$: a number of structures multiplied by simple factors ranging 
from $\cos^n\theta$ to $\sin^n \theta$ at the level of $n$-point functions. It would be very 
interesting to check whether such a simple form  arises for the $n$-point functions on the both 
sides of the duality.

\bigskip

\noindent {\bf Note added:} After the completion of our work, we learned that V.E. Didenko and M. A. Vasiliev are in the process of publishing results \cite{Didenko:2017lsn} on problems closely related to the subject matter of this paper.

%%%%%%%%%%%%%%%%%%%%%%%%%%%%%%%%%%%%%%%%%%%%%%%%%%%%%%%%%%%%%%%%%%%%%%%
\section*{Acknowledgments}
\label{sec:Aknowledgements}
%%%%%%%%%%%%%%%%%%%%%%%%%%%%%%%%%%%%%%%%%%%%%%%%%%%%%%%%%%%%%%%%%%%%%%%

We would like to thank Simone Giombi, Shiroman Prakash, Per Sundell and Sasha Zhiboedov for the very useful discussions and comments. E.Sk is grateful to Simone Giombi, V. Gurucharan, Volodya Kirilin and  Shiroman Prakash for the very useful discussions in the course of collaboration  \cite{Giombi:2016zwa}. The work 
of E.Se. and Y.Z. is supported in part by NSF grant PHY-1214344, in part by the Mitchell Institute 
for Fundamental Physics and Astronomy and in part by Conicyt MEC grant PAI 80160107. 
The work of E.Sk. was supported in part  by the Russian Science Foundation grant 14-42-00047 
in association with Lebedev Physical Institute and by the DFG Transregional Collaborative 
Research Centre TRR 33 and the DFG cluster of excellence ``Origin and Structure of the Universe". 
E.Se. thanks Pontifica Universidad Cat\'olica de Valparaiso for hospitality. E.Sk. thanks the 
Galileo Galilei Institute for Theoretical Physics (GGI) for the hospitality and INFN for partial 
support during the completion of this work, within the program “New Developments in $AdS_3/CFT_2$ Holography”.

\begin{appendix}

%%%%%%%%%%%%%%%%%%%%%%%%%%%%%%%%%%%%%%%%%%%%%%%%%%%%%%%%%%%%%%%%%%%%%%%
\section{ Notation and Conventions}
\label{app:notation}
%%%%%%%%%%%%%%%%%%%%%%%%%%%%%%%%%%%%%%%%%%%%%%%%%%%%%%%%%%%%%%%%%%%%%%%

We adopt the mostly plus convention for the metric $\eta_{\mm\nn} = (-+++)$, 
which makes the  Euclidean rotation easier to implement. 
Choosing $x^\mm=(\vec \rmx, z)$ 
to be Poincar\'e coordinates with $z$ being the radial coordinate and $\vec\rmx$ the three 
coordinates along the boundary, the $AdS_4$-background can be described by vierbein 
$h^{\ga\gad}_\mm$ and spin-connection that splits into (anti) self-dual parts 
$\omega_{(0)\mm}{}^{\ga\gb}$ and ${\bar\omega}_{(0)\mm}{}^{\gad\gbd}$:
\begin{align}
h^{\ga\gad} &=\frac1{2z} \sigma_\mm^{\ga\gad} dx^\mm\ , 
& \omega_{(0)}^{\ga\gb}& = \frac{i}{2z}\vec \sigma^{\ga\gb} \cdot d\vec x\ , &
{\bar\omega}_{(0)}^{\gad\gbd}& = -\frac{i}{2z}\vec\sigma^{\ga\gb}\cdot d\vec x\ . 
\end{align}
The matrices $\sigma_\mm^{\ga\gad}$ are constant and in our convention they are given by
$\sigma_\mm^{\ga\gbd}= (\vec \sigma^{\ga\gb}, i \epsilon^{\alpha\beta})$. We have the relations\footnote{The bulk 
objects may transform under the Lorentz algebra, $sl(2,\mathbb{C})$, while the choice of the Poincar\'e coordinates breaks the manifest symmetry down to the boundary Lorentz algebra, $sl(2,\mathbb{R})$. In particular, there exists $\epsilon^{\ga\gad}$ that allows one to map between dotted and undotted indices. For that reason the $3d$ coordinates $\rmx$ can carry different types of indices $\rmx^{\ga\gb}$, $\rmx^{\ga\gbd}$ or $\rmx^{\gad\gbd}$, while being the same object. }
\begin{align}
\rmx^{\ga\gb} &= \vec \sigma^{\ga\gb}\cdot  \vec \rmx\ ,  & 
{\rmx}^2 & = -\tfrac12 \rmx^{\ga\gb}\rmx_{\ga\gb}\ , &
\sigma^{\ga\gad}_\mm \sigma_{\nn\,\ga\gad} &=-2\eta_{\mm\nn}\ ,
\w2
 x^{\ga\gad} &=\rmx^{\ga\gad}+iz\epsilon^{\ga\gad}\ , &
x^2 &= \rmx^2 + z^2\ , &\rmx_{ij} &= |\rmx_i-\rmx_j|\ .
\end{align}
The inverse vierbein $h_{\ga\gad}^\mm = -z \sigma^\mm_{\ga\gad}$ obeys the relations 
\be
h^{\ga\gad}_\mm h^\nn_{\ga\gad}=\delta^\nn_\mm\ ,\qquad 
h^{\ga\gad}_\mm h^\mm_{\gb\gbd}=\delta_{\gb}^{\ga}\delta_{\gbd}^{\gad}\ , 
\ee
and the $AdS_4$ metric tensor and spin connection are given by
\bea
g_{\mm\nn} &=& -h^{\ga\gad}_\mm h\fd{\nn\,\ga\gad}dx^\mm dx^\nn 
=\frac1{2z^2} \eta_{\mm\nn}dx^\mm dx^\nn\ ,
\w2
\Omega &=& \frac{1}{4i}\left[~\omega_{(0)\,\alpha\beta}y^{\alpha}y^{\beta}+
\bar{\omega}_{(0)\dot\alpha\dot\beta} {\bar y}^{\,\dot\alpha}y^{\dot\beta}+
2 h_{\alpha\dot\beta}y^{\alpha}{\bar y}^{\,\dot\beta}\right]\ .
\eea
We use the raising and lowering conventions: $X^\alpha = \epsilon^{\alpha\beta} X_\beta$ and  
$X_\alpha = X^\beta \epsilon_{\beta\alpha}$ with 
$\epsilon_{\alpha\gamma}\epsilon^{\beta\gamma} =\delta_\alpha^\beta$, 
and similar conventions for the dotted indices. We use the convention  $\epsilon_{12}=-\epsilon_{21}=1$. It is sometimes convenient to define $\nabla_{\ga\gad}$ as $\nabla=\nabla_{\ga\gad} h^{\ga\gad}_\mm \, dx^\mm$.

%%%%%%%%%%%%%%%%%%%%%%%%%%%%%%%%%%%%%%%%%%%%%%%%%%%%%%%%%%%%%%%%%%%%%%%
\section{Triple Sums}
\label{app:sums}
%%%%%%%%%%%%%%%%%%%%%%%%%%%%%%%%%%%%%%%%%%%%%%%%%%%%%%%%%%%%%%%%%%%%%%%
Here we present some details on the triple sums that result from the bulk integrals.
%%%%%%%%%%%%%%%%%%%%%%%%%%%%%%%%%%%%%%%%%%%%%%%%%%%%%%%%%%%%%%%%%%%%%%%
\subsection{Type-A, Free Boson}
\label{app:freeboson}
%%%%%%%%%%%%%%%%%%%%%%%%%%%%%%%%%%%%%%%%%%%%%%%%%%%%%%%%%%%%%%%%%%%%%%%

We begin with the evaluation of the triple sum \eq{ts}. Expanding the last factor in \eq{ts}, we obtain 
\begin{align}
\sum_{l=0}^{s_2}\sum_{k=l}^{s_2}\sum_{i=0}^{k-l}\sum_{j=0}^{k-l-i}
&\frac{(-1)^{k+j+l}  
i^{s_1+s_2} 2^{s_1-s_2+1}   \Gamma (2 (-k+s_1+s_2))  
\Gamma \left(-i+k+s_1-s_2+\frac{1}{2}\right) }{\sqrt{\pi }  
i!j! l! (2 s_1-1)! (2 s_2-2k)!  (-i+2 s_1)!  (k-i-j-l)!}
\non\\ 
&\times \frac{Q_1^{s_1}Q_2^{s_2}\left(\frac{P^2_{12}}{Q_1Q_2}
\right)^{s_2-l}}{\rmx_{12}\rmx_{13}\rmx_{23}}\,.
\end{align}
After the first summation over $j$, the coefficient is proportional to ${}_1F_0(i-k+l,1)$. 
Since $k\geq i+l$, and ${}_1F_0(n,1)$ vanishes when $n$ is a negative integer, the second sum
over $i$ is equivalent to simply substituting $i\rightarrow k-l$. At this point we have
\begin{align}
\sum_{l=0}^{s_2}\sum_{k=l}^{s_2}&\frac{(-1)^{k+l}  i^{s_1+s_2} 2^{s_1-s_2+1}
\Gamma (2 (-k+s_1+s_2))  \Gamma \left(l+s_1-s_2+\frac{1}{2}\right) }{\sqrt{\pi } 
l! (2 s_1-1)! (2 s_2-2k)!  (-k+l+2 s_1)!  (k-l)!} 
\times \frac{Q_1^{s_1}Q_2^{s_2}\left(\frac{P^2_{12}}{Q_1Q_2}\right)^{s_2-l}}
{\rmx_{12}\rmx_{13}\rmx_{23}}\,.
\end{align}
We then evaluate the summation over $k$ and obtain
\begin{align}
\label{fbsum}
&\qquad\qquad\sum_{l=0}^{s_2}\frac{i^{s_1+s_2} 2^{s_1-s_2+1} \Gamma (2 (-l+s_1+s_2))  
\Gamma \left(l+s_1-s_2+\frac{1}{2}\right) }{\sqrt{\pi } l! (2 s_1-1)!(2s_1)!(2s_2-2l)!}
\non\\
&\times{}_3F_2(-2s_1,l-s_2,\frac{1}{2}+l-s_2;\frac{1}{2}+l-s_1-s_2,1+l-s_1-s_2;1)\times \frac{Q_1^{s_1}Q_2^{s_2}\left(\frac{P^2_{12}}{Q_1Q_2}\right)^{s_2
-l}}{\rmx_{12}\rmx_{13}\rmx_{23}}\,.
\end{align}
Using the identity
\be
_3F_2(-n,b,c;d,b+c-d-n+1;1)=\frac{(d-b)_n(d-c)_n}{(d)_n(d-b-c)_n}~,~~n\in\mathbb{N}\, ,
\ee
where $(a)_n=a(a+1)\cdots(a+n-1)$ is the Pochhammer symbol, the hypergeometric function 
can then be written in terms of gamma functions
\begin{align}
&_3F_2(-2s_1,l-s_2,\frac{1}{2}+l-s_2;\frac{1}{2}+l-s_1-s_2,1+l-s_1-s_2;1)
\non\\
=&\frac{(\frac{1}{2}-s_1)_{2s_1}(-s_1)_{2s_1}}{(\frac{1}{2}+l-s_1-s_2)_{2s_1}(-l-s_1+s_2)_{2s_1}}=\frac{\Gamma(2s_1)\Gamma(1+2s_1)}{\Gamma(1+2l+2s_1-2s_2)\Gamma(2(-l+s_1+s_2))}~.
\end{align}
Note that $s_1\geq s_2\geq l$, and a cancellation of zeros in $(-s)_{2s_1}$ and $(-l-s_1+s_2)_{2s_1}$ 
has been performed to get the bottom line. Replacing the hypergeometric function by the 
above expression, one can see that 
the coefficients in \eqref{fbsum} are equal to the ones from the CFT side, c.f. \eqref{bosonjjoU}.

%%%%%%%%%%%%%%%%%%%%%%%%%%%%%%%%%%%%%%%%%%%%%%%%%%%%%%%%%%%%%%%%%%%%%%%
\subsection{Type-B, Free Fermion}
\label{app:freefermion}

We now consider the triple sum \eq{ts2}, which upon the expansion of the last factor, can be written as 
\begin{align}
\sum_{l=0}^{s_2-1}\sum_{k=l}^{s_2-1}\sum_{i=0}^{k-l}\sum_{j=0}^{k-l-i}&\frac{(-1)^{k+j+l}  
i^{s_1+s_2} 2^{s_1-s_2+1}   \Gamma (2 (-k+s_1+s_2)-1)  
\Gamma \left(-i+k+s_1-s_2+\frac{3}{2}\right) }{\sqrt{\pi }  
i!j! l! (2 s_1-1)! (2 s_2-2k-1)!  (-i+2 s_1)!  (k-i-j-l)!}\non
\\ 
&\times \frac{S_3Q_1^{s_1}Q_2^{s_2}\left(\frac{P^2_{12}}{Q_1Q_2}
\right)^{s_2-l}(P_{12})^{-1}}{\rmx^2_{23}\rmx^2_{13}}\,.
\end{align}
As in the type-A model case, after the first summation over $j$, the coefficient of each term 
is proportional to ${}_1F_0(i-k+l,1)$, therefore for the same reason the second summation 
over $i$ is equivalent to the substitution $i\rightarrow k-l$ in the summand. The result is then
\begin{align}
\sum_{l=0}^{s_2-1}\sum_{k=l}^{s_2-1}&\frac{(-1)^{k+l}  i^{s_1+s_2} 2^{s_1-s_2+1} 
\Gamma (2 (-k+s_1+s_2)-1)  \Gamma \left(l+s_1-s_2+\frac{3}{2}\right) }{\sqrt{\pi } 
l! (2 s_1-1)! (2 s_2-2k-1)!  (-k+l+2 s_1)!  (k-l)!}
\non\\ 
&\times \frac{S_3Q_1^{s_1}Q_2^{s_2}\left(\frac{P^2_{12}}{Q_1Q_2}
\right)^{s_2-l}(P_{12})^{-1}}{\rmx^2_{23}\rmx^2_{13}}\,.
\end{align}
We sum over $k$ and obtain
\begin{align}
\label{ffsum}
&\qquad\sum_{l=0}^{s_2-1}\frac{i^{s_1+s_2} 2^{s_1-s_2+1} \Gamma (2 (-l+s_1+s_2)-1)  
\Gamma \left(l+s_1-s_2+\frac{3}{2}\right) }{\sqrt{\pi } l! (2 s_1-1)!(2s_1)!(2s_2-2l-1)!}
\non\\
&\times{}_3F_2(-2s_1,\frac{1}{2}+l-s_2,1+l-s_2;1+l-s_1-s_2,\frac{3}{2}+l-s_1-s_2;1)
\non\\ 
&\times \frac{S_3Q_1^{s_1}Q_2^{s_2}\left(\frac{P^2_{12}}{Q_1Q_2}
\right)^{s_2-l}(P_{12})^{-1}}{\rmx^2_{23}\rmx^2_{13}}\,.
\end{align}
Using the fact that the hypergeometric function in the above expression equals 
\be
\frac{\Gamma(2s_1)\Gamma(2s_1+1)}{\Gamma(2s_1-2s_2+2l+2)\Gamma(2s_1+2s_2-2l-1)}\ ,
\ee
we find that the coefficient of each term in \eqref{ffsum} matches the corresponding one from CFT three-point function exactly, c.f. \eqref{freefermjooU}.

%%%%%%%%%%%%%%%%%%%%%%%%%%%%%%%%%%%%%%%%%%%%%%%%%%%%%%%%%%%%%%%%%%%%%%%

%%%%%%%%%%%%%%%%%%%%%%%%%%%%%%%%%%%%%%%%%%%%%%%%%%%%%%%%%%%%%%%%%%%%%%%
\subsection{Type-A, Critical Boson}
\label{app:criticalboson}
%%%%%%%%%%%%%%%%%%%%%%%%%%%%%%%%%%%%%%%%%%%%%%%%%%%%%%%%%%%%%%%%%%%%%%%

Next, we evaluate the summations in \eqref{tAd2sum}. This can be done in an easy way by 
keeping the invariant structure $S_3$ instead of $Q_2$:
\begin{align}
\frac{\tilde{c}_0}{c_0}\sum_{k=0}^{s_2}\sum_{i=0}^{k}
\sum_{j=0}^{k-i}&\frac{(-1)^{k-j} 2^{s_1-s_2+1} i^{s_1+s_2} 
\Gamma (-2 k+2 s_1+2 s_2) \Gamma (-i+k+s_1-s_2+1)}{\sqrt{\pi } i! j! 
\Gamma (2 s_1) \Gamma (-i+2 s_1+1) \Gamma (-2 k+2 s_2+1) \Gamma 
\left(-i-j+k+\frac{1}{2}\right)}
\non
\\ 
&\times \frac{(Q_1)^{s_1-s_2} 
(P_{12})^{2(i+j-k+s_2)} \left((S_3)^2\right)^{-i-j+k}}{\rmx^2_{23}\rmx_{13}^2}\ .
\end{align}
For the $(S_3)^2$ term with power $l=k-i-j$, we have
\begin{align}
\frac{\tilde{c}_0}{c_0}\sum_{l=0}^{s_2}\sum_{k=l}^{s_2}
\sum_{j=0}^{k-l}&\frac{(-1)^{k-j} 2^{s_1-s_2+1} i^{s_1+s_2} 
\Gamma (-2 k+2 s_1+2 s_2) \Gamma (l+j+s_1-s_2+1)}{\sqrt{\pi } (k-l-j)! j!
\Gamma (2 s_1) \Gamma (-k+l+j+2 s_1+1) \Gamma (-2 k+2 s_2+1) \Gamma 
\left(l+\frac{1}{2}\right)}
\non
\\ 
&\times \frac{(Q_1)^{s_1-s_2} 
(P_{12})^{2 (s_2-l)} (S_3)^{2l}}{\rmx^2_{23}\rmx_{13}^2}\ .
\end{align}
Notice that the summation over $i$ is now replaced by $i\rightarrow k-l-j$. After the first summation we obtain
\begin{align}
\frac{\tilde{c}_0}{c_0}\sum_{k=l}^{s_2}&\frac{(-1)^ki^{s_1+s_2}2^{1+s_1-s_2}
\Gamma(1+l+s_1-s_2)\Gamma\left(
2(-k+s_1+s_2)\right)}{\sqrt{\pi}\Gamma(1+k-l)\Gamma(\frac{1}{2}+l) \Gamma(2s_1)
\Gamma(1-2k+2s_2)\Gamma(1-k+l+2s_1)} 
\non\\
&\times {}_2F_1(-k+l,1+l+s_1-s_2,1-k+l+2s_1,1) \times \frac{(Q_1)^{s_1-s_2} 
(P_{12})^{2 (s_2-l)} (S_3)^{2l}}{\rmx^2_{23}\rmx_{13}^2}\ .
\end{align}
Replacing $_2F_1(-k+l,1+l+s_1-s_2,1-k+l+2s_1,1)$ by
%e
\be
\frac{\Gamma(1-k+l+2s_1)\Gamma(s_1+s_2-l)}{\Gamma(1+2s_1)\Gamma(s_1+s_2-k)}\ ,
\ee
we can evaluate the second summation and the result is
\begin{align}
\frac{\tilde{c}_0}{c_0}\sum_{l=0}^{s_2}&\frac{(-1)^li^{s_1 + s_2}2^{1+s_1-s_2}
\Gamma(1+l+s_1-s_2)\Gamma\left(2(-l+s_1+s_2)\right)}{\sqrt{\pi}
\Gamma(\frac{1}{2}+l)\Gamma(2s_1) \Gamma(1+2s_1)\Gamma(1-2l+2s_2)} 
\non\\
&\times {}_2F_1(l-s_2,\frac{1}{2}+l-s_2,\frac{1}{2}+l-s_1-s_2,1) 
\non\\
&\times \frac{(Q_1)^{s_1-s_2} 
(P_{12})^{2 (s_2-l)} (S_3)^{2l}}{\rmx^2_{23}\rmx_{13}^2}~~~~(\text{Type-A,}~\Delta=2)\ .
\end{align}
This formula explicitly gives the Taylor coefficients of the generating function for critical 
boson theory. To facilitate the comparison between Type-A models with different boundary 
conditions, we rewrite \eqref{ts} in terms of $Q_1, P_{12}$ and $S_3$, and redo the 
summations, following the above procedure. The answer is
\begin{align}
\sum_{l=0}^{s_2}&\frac{(-1)^li^{s_1 + s_2}2^{1+s_1-s_2}\Gamma(\frac{1}{2}+l+s_1-s_2)
\Gamma\left(2(-l+s_1+s_2)\right)}{\sqrt{\pi}
\Gamma(1+l)\Gamma(2s_1) \Gamma(1+2s_1)\Gamma(1-2l+2s_2)} 
\non\\
&\times {}_2F_1(l-s_2,\frac{1}{2}+l-s_2,1+l-s_1-s_2,1) 
\non\\
&\times \frac{(Q_1)^{s_1-s_2} 
(P_{12})^{2 (s_2-l)} (S_3)^{2l}}{\rmx_{12}\rmx_{13}\rmx_{23}}~~~~(\text{Type-A,}~\Delta=1)\ .
\end{align}
This is just \eqref{fbsum} but written in terms of $Q_1, P_{12}$ and $S_3$ instead.

%%%%%%%%%%%%%%%%%%%%%%%%%%%%%%%%%%%%%%%%%%%%%%%%%%%%%%%%%%%%%%%%%%%%%%%
\subsection{Type-B, Critical Fermion}
\label{app:criticalfermion}
%%%%%%%%%%%%%%%%%%%%%%%%%%%%%%%%%%%%%%%%%%%%%%%%%%%%%%%%%%%%%%%%%%%%%%%

Starting with \eqref{tbd1sum}, we use invariant structure $S_3$ and make the change of the index $i\rightarrow k-l-j$:
\begin{align}
\sum_{l=0}^{s_2-1}\sum_{k=l}^{s_2-1}\sum_{j=0}^{k-l}&\frac{(-1)^{j+k} i^{s_1+s_2} 2^{s_1-s_2+1} 
\Gamma (-2 k+2 s_1+2 s_2-1) \Gamma (l+j+s_1-s_2+1)}{\sqrt{\pi } (k-l-j)! j! 
\Gamma (2 s_1) (-2 k+2 s_2-1)! \Gamma (l+j-k+2s_1+1) \Gamma \left(l+\frac{3}{2}
\right) }\non
\\ 
&\times \frac{(Q_1)^{s_1-s_2} (P_{12})^{2(s_
2-l)-1} ( S_3)^{2l+1}}{\rmx_{12}\rmx_{13}\rmx_{23}}\ .
\end{align}
As in the case of Type-A with $\Delta=2$, it is straightforward to evaluate the first two sums. 
The result is
\begin{align}
\sum_{l=0}^{s_2-1}&\frac{(-1)^li^{s_1 + s_2}2^{s_1-s_2+1}\Gamma(1+l+s_1-s_2)
\Gamma(2s_1+2s_2-2l-1)}{\sqrt{\pi}
\Gamma(\frac{3}{2}+l)\Gamma(2s_1)\Gamma(2s_1+1)\Gamma(2s_2-2l)} 
\non\\
&\times {}_2F_1(\frac{1}{2}+l-s_2,1+l-s_2,\frac{3}{2}+l-s_1-s_2,1) \Big)
\non\\
&\times \frac{(Q_1)^{s_1-s_2} (P_{12})^{2(s_
2-l)-1} ( S_3)^{2l+1}}{\rmx_{12}\rmx_{13}\rmx_{23}}~~~~(\text{Type-B,}~\Delta=1)\ .
\end{align}
One can also rewrite \eqref{ffsum} in terms of $Q_1, P_{12}$ and $S_3$, and the result is
\begin{align}
\sum_{l=0}^{s_2-1}&\frac{(-1)^li^{s_1 + s_2}2^{s_1-s_2+1}
\Gamma(\frac{3}{2}+l+s_1-s_2)\Gamma(2s_1+2s_2-2l-1)}{\sqrt{\pi}
\Gamma(1+l)\Gamma(2s_1)\Gamma(2s_1+1)\Gamma(2s_2-2l)} 
\non\\
&\times {}_2F_1(\frac{1}{2}+l-s_2,1+l-s_2,1+l-s_1-s_2,1)
\non\\
&\times \frac{(Q_1)^{s_1-s_2} (P_{12})^{2(s_
2-l)-1} ( S_3)^{2l+1}}{\rmx_{23}^2\rmx_{13}^2}~~~~(\text{Type-B,}~\Delta=2)\ .
\end{align}
%

%%%%%%%%%%%%%%%%%%%%%%%%%%%%%%%%%%%%%%%%%%%%%%%%%%%%%%%%%%%%%%%%%%%%%%%

\end{appendix}

\newpage

%\setstretch{0.90}
\providecommand{\href}[2]{#2}\begingroup\raggedright\endgroup

\end{document}